\documentclass[preprintnumbers, floatfix,letterpaper,aps,prd,epsfig,nofootinbib,
twocolumn
]{revtex4-1}

\usepackage{bm,graphicx,dcolumn,epstopdf,epsf, latexsym,mathbbol, amssymb,amsmath,color,slashed, mathrsfs,mathcomp,simplewick}
\usepackage[center]{subfigure}
\usepackage{multirow}
\usepackage{makecell}
\usepackage[colorlinks,linkcolor=blue,citecolor=blue,urlcolor=blue]{hyperref}
\begin{document}
\renewcommand\arraystretch{2}
 \newcommand{\bq}{\begin{equation}\begin{aligned}[b]}
 \newcommand{\eq}{\end{aligned}\end{equation}}
 \newcommand{\bqn}{\begin{eqnarray}}
 \newcommand{\eqn}{\end{eqnarray}}
 \newcommand{\nb}{\nonumber}
 \newcommand{\bvec}[1]{\mathbf #1}
 \newcommand{\cb}{\color{blue}}
    \newcommand{\cc}{\color{cyan}}
     \newcommand{\lb}{\label}
        \newcommand{\cm}{\color{magenta}}
\newcommand{\rc}{\rho^{\scriptscriptstyle{\mathrm{I}}}_c}
\newcommand{\rd}{\rho^{\scriptscriptstyle{\mathrm{II}}}_c}
%\NewDocumentCommand{\evalat}{sO{\big}mm}{%
%  \IfBooleanTF{#1}
%   {\mleft. #3 \mright|_{#4}}
%   {#3#2|_{#4}}
%}

\newcommand{\PRL}{Phys. Rev. Lett.}
\newcommand{\PL}{Phys. Lett.}
\newcommand{\PR}{Phys. Rev.}
\newcommand{\CQG}{Class. Quantum Grav.}
\newcommand{\parallelsum}{\mathbin{\!/\mkern-5mu/\!}}

\title{Forecasts for Constraining Lorentz-violating Damping of Gravitational Waves from Compact Binary Inspirals}

\author{Bo-Yang Zhang${}^{a, b, c}$}
\email{zhangby@stumail.neu.edu.cn}

\author{Tao Zhu${}^{b, c}$}
\email{Corresponding author: zhut05@zjut.edu.cn}

\author{Jing-Fei Zhang${}^{a}$}
\email{Corresponding author: jfzhang@mail.neu.edu.cn}

\author{Xin Zhang${}^{a,d,e}$}
\email{zhangxin@mail.neu.edu.cn}

\affiliation{
%${}^{a}$ Key Laboratory of Cosmology and Astrophysics (Liaoning) \& Department of Physics, College of Sciences, Northeastern University, Shenyang 110819, China\\
${}^{a}$ Key Laboratory of Cosmology and Astrophysics (Liaoning) \& College of Sciences, Northeastern University, Shenyang 110819, China\\
${}^{b}$ Institute for Theoretical Physics and Cosmology, Zhejiang University of Technology, Hangzhou, 310032, China\\
${}^{c}$ United Center for Gravitational Wave Physics (UCGWP), Zhejiang University of Technology, Hangzhou, 310032, China\\
${}^{d}$ Key Laboratory of Data Analytics and Optimization for Smart Industry (Ministry of Education), Northeastern University, Shenyang 110819, China\\
${}^{e}$ National Frontiers Science Center for Industrial Intelligence and Systems Optimization, Northeastern University, Shenyang 110819, China}

\begin{abstract}

Violation of Lorentz symmetry can result in two distinct effects in the propagation of the gravitational waves (GWs). One is a modified dispersion relation and another is a frequency-dependent damping of GWs. While the former has been extensively studied in the literature, in this paper we concentrate on the frequency-dependent damping effect that arises from several specific Lorentz-violating theories, such as spatial covariant gravities, Ho\v{r}ava-Lifshitz gravities, etc. This Lorentz-violating damping effect changes the damping rate of GWs at different frequencies and leads to an amplitude correction to the GW waveform of compact binary inspiral systems. With this modified waveform, we then use the Fisher information matrix to investigate the prospects of constraining the Lorentz-violating damping effect with GW observations. We consider both ground-based and space-based GW detectors, including the advanced LIGO, Einstein Telescope, Cosmic Explorer (CE), Taiji, TianQin, and LISA. Our results indicate that the ground-based detectors in general give tighter constraints than those from the space-based detectors. Among the considered three ground-based detectors, CE can give the tightest constraints on the Lorentz-violating damping effect, which improves the current constraint from LIGO-Virgo-KAGRA events by about 8 times.

\end{abstract}

\maketitle

\section{INTRODUCTION}
\renewcommand{\theequation}{1.\arabic{equation}} \setcounter{equation}{0}

Since the landmark discovery of the first gravitational wave (GW) event, GW150914, resulting from the coalescence of two massive black holes by the LIGO-Virgo collaboration in 2015 \cite{LIGOScientific:2016aoc}, the field of GW astronomy has rapidly evolved.
To date, approximately 90 events have been meticulously identified by the LIGO-Virgo-KAGRA (LVK) scientific collaborations \cite{LIGOScientific:2018mvr, LIGOScientific:2019lzm, LIGOScientific:2020aai, LIGOScientific:2021djp}. On May 24, 2023, the advanced LIGO (aLIGO) initiated the Observing Run 4 (O4) project. Following LVK, the forthcoming third-generation ground-based GW detectors, such as Einstein Telescope (ET) \cite{Branchesi:2023mws} and Cosmic Explorer (CE) \cite{Evans:2021gyd}, are presently in the design phase, with a specific emphasis on detecting high-redshift GW events ($z > 10$). Concurrently, a new cohort of space-based detectors (Taiji \cite{Ruan:2018tsw, Wu:2018clg, Hu:2017mde}, TianQin \cite{Liu:2020eko, Wang:2019ryf, TianQin:2015yph, Luo:2020bls, Milyukov:2020kyg}, and LISA \cite{Robson:2018ifk, LISACosmologyWorkingGroup:2022jok}) is designed to explore the low-frequency GW signals ($f\sim 10^{-4}$ Hz). We anticipate that these detectors will play a crucial role in the era of GW astronomy \cite{LIGOScientific:2016lio, LIGOScientific:2018dkp, LIGOScientific:2017zic, LIGOScientific:2019fpa, LIGOScientific:2020tif}. 

General relativity (GR) remains the preeminent theory for explaining gravitational phenomena. Yet, it faces challenges in accounting for enigmatic concepts such as dark matter and dark energy, and reconciling them with quantum mechanics, particularly in the contexts of singularities and the quantization of gravity. To address these issues, a plethora of experiments have been devised to rigorously test GR's predictions. Regrettably, the majority of these experiments have been limited to investigating the weak-field regime \cite{Hoyle:2000cv, Everitt:2011hp, Sabulsky:2018jma}. GWs, one of the fundamental predictions of GR, are produced in the tumultuous environments of strong gravitational fields and interact only weakly with matter, making them pristine messengers of the dynamics of space-time. The detection of GWs, especially those originating from the coalescences of compact binary systems \cite{LIGOScientific:2020iuh,  LIGOScientific:2017vwq, LIGOScientific:2020ibl}, has thus heralded a new era for testing the robustness of GR under extreme conditions. These observations offer a powerful tool for probing the strong-field regime of gravity, potentially unlocking answers to the persistent questions that challenge the current understanding within the framework of GR.

In the theoretical realm, various modified theories of gravity have been proposed to address challenges within GR, see refs.~\cite{Cognola:2006eg, Copeland:2006wr, Frieman:2008sn, Jackiw:2003pm, Yunes:2010yf, Yagi:2012vf, Wang:2017brl, Sefiedgar:2010we, Crisostomi:2017ugk, Gao:2014soa, Gao:2019lpz} and references therein. A subset of these theories has garnered significant attention for deviating from a fundamental principle of GR --- the Lorentz invariance. At high-energy levels, it is widely believed that this invariance will be broken when gravity is quantized. Various modified gravity theories have been proposed to explore the nature of Lorentz violation in gravity, including the Einstein-\AE{}ther theory \cite{Jacobson:2000xp, Eling:2004dk, Jacobson:2007veq, Li:2007vz, Battye:2017zvv, Zhang:2023kzs, Liu:2021yev, Zhu:2019ura}, Ho\v{r}ava-Lifshitz theories of quantum gravity \cite{Horava:2009uw, Takahashi:2009wc,  Wang:2012fi, Zhu:2013fja}, and spatial covariant gravities \cite{Gao:2020qxy, Gao:2020yzr, Gao:2019twq, Gao:2019liu, Joshi:2021azw}. A phenomenological framework, the  standard model extension, has also been extensively studied in the literature for exploring the possible properties of Lorentz violations in the gravitational sector \cite{Kostelecky:2016kfm, Bailey:2006fd, Mewes:2019dhj, Shao:2020shv, Wang:2021ctl, Kostelecky:2017zob, Niu:2022yhr}.

The violation of Lorentz symmetry in gravity can introduce deviations from GR in the propagation of GWs. These deviations manifest in two distinct ways, influencing the propagation behavior of GWs in the cosmological background. First, with Lorentz violation, the conventional linear dispersion relation of GWs can be modified into a nonlinear one, which in turn changes the phase velocities of GWs at different frequencies. This effect can arise from a large number of Lorentz-violating theories. Second, Lorentz violation can introduce frequency-dependent friction into the propagation equation of GWs, resulting in varying damping rates for GWs of different frequencies during their propagation. This effect normally arises from those theories with mixed temporal and spatial derivatives of the spacetime metric in the modified theories of gravity with spatial covariance, for example, the Ho\v{r}ava-Lifshitz gravity \cite{Colombo:2014lta}, the spatial covariant gravities \cite{Gao:2019liu, Zhu:2022uoq}, etc.  Here we would like to note that the possible Lorentz violations could also lead to source-dependence on the speed of GWs \cite{Ghosh:2023xes}. 

Testing Lorentz symmetry of gravitational interaction by using the observational data from GW events in LIGO-Virgo-KAGRA catalogs and future GW detectors has been carried out in a lot of works, see Refs. \cite{LIGOScientific:2019fpa, LIGOScientific:2020tif, Mirshekari:2011yq, Will:1997bb, Zhu:2022uoq, Gong:2021jgg, Haegel:2022ymk} and references therein. In most of these works, the effects due to the nonlinear dispersion relation have been extensively considered. Recently, the constraint on the Lorentz-violating damping effects from GW events in LIGO-Virgo-KAGRA catalogs was first obtained \cite{Zhu:2023wci}. 

In this paper, we detail the Lorentz-violating damping effects in the propagation of GW in a cosmological background. Decomposing the GWs into left-hand and right-hand circular polarization modes, we observe that the Lorentz-violating damping effects manifest through explicit modifications in the GW amplitude. We derive corrections to the waveform of the compact binary inspiral system accordingly. With this modified waveform, we use the Fisher information matrix (FIM), which is widely used in cosmology and astrophysics \cite{Yu:2023ico, Jin:2023tou, Shao:2023agv, Jin:2023sfc, Zhang:2023gaz, Wu:2022jkf, Song:2022siz, Jin:2022qnj, Wu:2022dgy, Wang:2022oou, Wu:2021vfz, Jin:2021pcv, Zhang:2021yof,  Yu:2021nvx, Zhao:2019gyk, Zhang:2019loq, Wang:2019tto, Zhang:2019ple} to investigate the prospects of constraining the Lorentz-violating damping effect with GW observations of compact binary systems. We consider both ground-based and space-based GW detectors, including aLIGO, CE, ET, Taiji, TianQin, and LISA. Our results indicate that the ground-based detectors in general give tighter constraints than those from the space-based detectors. Among the considered three ground-based detectors, CE can give the tightest constraints on the Lorentz-violating damping effect, which improves the current constraint from LIGO-Virgo-KAGRA events by about 8 times \cite{Zhu:2023wci}.

Our paper is organized as follows. In Sec. \ref{sec:2}, We present a very brief introduction to the Lorentz-violating damping effect and calculate the modified waveform of GWs of compact binary inspiral systems with the effect. Sec. \ref{sec:3} summarizes the application of the FIM for constraining the modified waveform parameters of GWs. The main results of our analysis are discussed in Sec. \ref{sec:4}. Finally, Sec. \ref{sec:5} provides a summary and further discussion of our work in this paper.

\section{MODIFIED WAVEFORM OF GWS WITH LORENTZ-VIOLATING DAMPING EFFECT}\label{sec:2}
\renewcommand{\theequation}{2.\arabic{equation}} \setcounter{equation}{0}

In this section, we present a brief introduction to the modified waveform of GWs with Lorentz-violating damping effect. As we mentioned, such Lorentz-violating damping effect can modify the amplitude damping rates of the two tensorial modes of GWs, which arise from several specific Lorentz-violating theories of gravity, for instance, the spatial covariant gravities \cite{Gao:2019liu, Zhu:2022uoq} and Ho\v{r}ava-Lifshitz gravity \cite{Colombo:2014lta}. 

\begin{figure*}
\includegraphics[width=14cm]{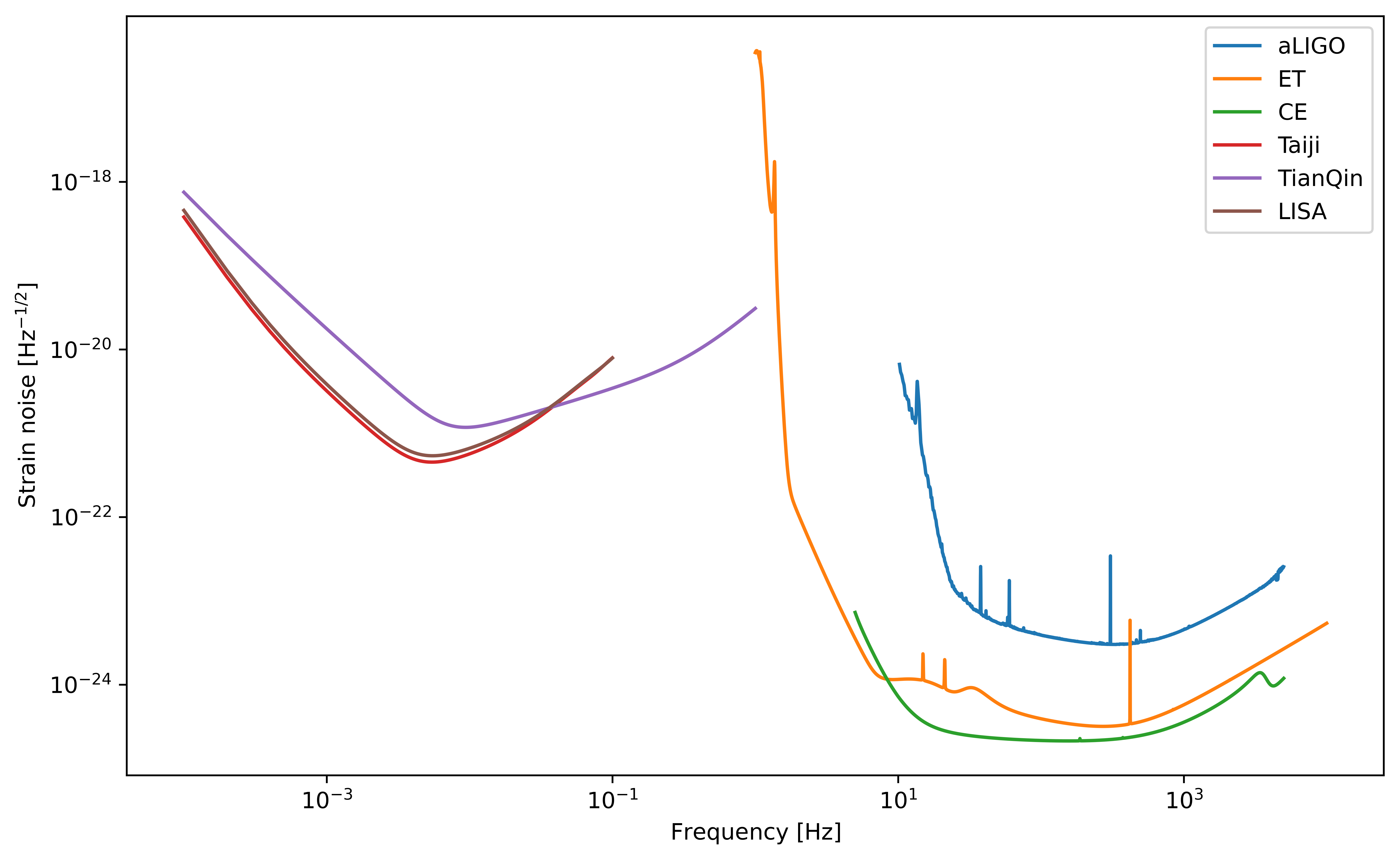}
\caption{\label{fig:detectors} The noise spectral density of the six detectors considered in this paper. Both ground-based and space-based detectors are included in the picture.}
\end{figure*}

\subsection{Propagating equation of GWs with Lorentz-violating damping effects}

Let us investigate the propagation of GWs with Lorentz-violating damping effect on a flat Friedmann-Robertson-Walker spacetime. Treating this spacetime as a background, GWs can be described by the tensor perturbations of the metric, where the metric is expressed in the form of
\begin{eqnarray}
ds^2= a^2(\tau) \Big[-d\tau^2+ (\delta_{ij}+h_{ij})dx^idx^j\Big],
\end{eqnarray}
where $a(\tau)$ is the scale factor of the expanding Universe and $\tau$ represents the conformal time. One can transform the conformal time $\tau$ to the cosmic time $t$ by $dt=a(\tau)d\tau$. Throughout this paper, we set the present expansion factor $a_0=1$. $h_{ij}$ denote GWs, which are transverse and traceless, i.e.,
\begin{eqnarray}
    \partial^i h_{ij} = 0 = h^i_i.
\end{eqnarray}
For later convenience, let us expand $h_{ij}$ over spatial Fourier harmonics,
\begin{eqnarray}
    h_{ij}(\tau, x^i) = \sum_{A=R, L} \int \frac{d^3k}{(2\pi)^3}h_A(\tau, k^i) e^{i k_i x^i} e_{ij}^A(k^i),\label{Fourier}
\end{eqnarray}
where $e_{ij}^A$ is the circular polarization tensor and obeys the following rules
\begin{eqnarray}
  \epsilon^{ijk}n_i e^A_{kl} = i \rho_A e_l^{j A}
\end{eqnarray}
with $\rho_{\rm R}=1$ and $\rho_{\rm L}=-1$. 

To study the Lorentz-violating damping effect on the propagation of GWs, let us first write the modified propagation equation of motions of the two GW modes in the following parametrized form \cite{Zhu:2023wci, Zhao:2019xmm},
\begin{eqnarray}
h_{A}^{\prime \prime} + (2 + {\bar \nu} + \nu_{A}) \mathcal{H} h^{\prime}_{A} + (1 + \bar{\mu} + \mu_{A}) k^{2} h_{A} = 0, \label{parametrized_eq} \nb\\
\end{eqnarray}
where a prime denotes the derivative concerning the conformal time $\tau$ and ${\cal H}=a'/a$. The four parameters, ${\bar \nu}$, $\nu_A$, ${\bar \mu}$, and $\mu_A$ label the new effects on the propagation of GWs arising from theories beyond GR. As mentioned in Ref. \cite{Zhu:2023wci}, such parametrization provides a general framework for exploring possible modified GW propagations arising from a large number of modified theories of gravity. Different parameters correspond to different effects on the propagation of GWs. The parameters $\nu_A$ and $\mu_A$ represent the effects of parity violations, while the parameters ${\bar \nu}$ and ${\bar \mu}$, if frequency-dependent, can originate from other potential modifications involving Lorentz violations. For ${\bar \nu}$ and ${\bar \mu}$, the former provides an amplitude modulation of the GW waveform, while the latter one ${\bar \mu}$ determines the phase velocities of the GWs.

In this paper, we will only concentrate on the case of the Lorentz-violating damping effect, and for this case one has
\begin{eqnarray}
{\bar \nu} \neq 0, \;\; \nu_A=0, \;\; {\bar \mu}=0=\mu_A.
\end{eqnarray}
In general, due to whether the parameter $\bar \nu$ is frequency-independent or not, the effects of $\bar \nu$ have two possibilities. When $\bar \nu$ is frequency-independent and time-dependent, it can be related to a time-dependent Planck mass $M_*(t)$ by writing \cite{Lagos:2019kds}
\bqn
{\cal H} {\bar \nu} = H \frac{d \ln M_*^2}{\ln a}. 
\eqn
See ref.~\cite{Zhu:2022uoq} as well for a specific example with an explicit action for nonzero $\bar \nu$ and its relation to the running of the Planck mass $M_*(t)$. Another possibility corresponds to a frequency-dependent $\bar \nu$, which represents the Lorentz-violating damping effect we studied in this paper. For this case, following Ref. \cite{Zhu:2023wci}, one can further parametrize ${\bar \nu}$ in the form of
\begin{eqnarray}
{\cal H} {\bar \nu} = \left[\alpha_{\bar \nu}(\tau) \left(\frac{k}{a M_{\rm LV}}\right)^{\beta_{\bar \nu}}\right]',
\end{eqnarray}
where $\beta_{\bar \nu}$ is an arbitrary number, $\alpha_{\bar \nu}$ is an arbitrary function of time, and $M_{\rm LV}$ denotes the energy scale of the Lorentz violation \footnote{In ref.~\cite{Zhao:2019xmm}, a different symbol $M_{\rm PV}$ is used to represent the Lorentz-violating energy scale. Note that $M_{\rm LV}^{-2}$ in the above parametrization is also directly related to the coefficient ${\cal G}_2/{\cal G}_0$ used for parametrizing the modified GW propagarions in ref.~\cite{Gao:2019liu}. }. The parameters $\alpha_{\bar \nu}$ and $\beta_{\bar \nu}$ depend on the specific modified theories of gravity. This case can arise from the mixed temporal and spatial derivatives of the spacetime metric in the modified theories of gravity with spatial covariance \cite{Gao:2019liu, Zhu:2022uoq, Colombo:2014lta}. In the next subsection, we present a specific example that induces the Lorentz-violating damping effect in the propagation of GWs.

\subsection{A specific example with Lorentz-violating damping effect}

To illustrate the Lorentz-violating damping effects clearly, let us consider a specific example, with the mixed term $ \nabla_k K_{ij} \nabla^k K^{ij}$, which can appear in both the spatial covariant gravity \cite{Gao:2019liu} and Ho\v{r}ava-Lifshitz gravity \cite{Colombo:2014lta}, where $\nabla_k$ denotes the covariant derivative associated with the spatial metric $g_{ij}$ and $K_{ij}$ is the extrinsic curvature tensor. It is also shown that by including mixed derivative terms, the non-protectable Ho\v{r}ava-Lifshitz gravity could be power-counting renormalizable and free of ghosts \cite{Colombo:2014lta}. With this mixed term, one can write down the general action of the gravitational part with spatial covariance in the form of \cite{Zhu:2022uoq}
\begin{eqnarray}
S &=& \frac{M_{\mathrm{Pl}}^{2}}{2} \int dtd^{3}x \sqrt{g} N (K_{ij} K^{ij} + R - K^{2}) \nonumber \\
&&+ \frac{M_{\mathrm{Pl}}^{2}}{2} \int dtd^{3}x \sqrt{g} N c_{1} (\nabla_{k}K_{ij}\nabla^{k}K^{ij} - R_{ij}R^{ij}), \nonumber \\ \lb{action_ADM}
\end{eqnarray}
where the first term represents the Einstein-Hilbert action of GR in the $3+1$ form, the second term signifies one of the modifications to GR, $c_{1}$ is the coupling coefficient which is a function of the lapse function $N$ and time $t$, and $M_{\rm Pl}$ is the reduced Planck mass. Here we would like to mention that in the second term of the action, we also include $R_{ij}R^{ij}$ to eliminate the effect of the mixed derivative term $\nabla_k K_{ij} \nabla^k K^{ij}$ in the disperation of GWs, such that the GWs propagate at the speed of light. Note that in writting the above action, we adopt the Arnowitt-Deser-Misner (ADM) form \cite{ADM}. In eq.(\ref{action_ADM}), $N$ is the lapse function, $\nabla_k$ denotes the covariant derivative associated with the spatial metric $g_{ij}$ and $K_{ij}$ is the extrinsic curvature tensor in the ADM form. Then, the action of GWs with the $c_{1}$ term up to the quadratic order can be written in the form \cite{Gao:2019liu},
\begin{eqnarray}\label{quadratic}
S^{(2)} &=& \frac{M_{\rm Pl}^2}{8}\int dtd^{3}x a^{3} \left( \dot{h}_{ij}\dot{h}^{ij} +{h}_{ij}\frac{\triangle}{a^{2}}{h}^{ij} \right. \nb\\
&&  \left.- c_1 \dot{h}_{ij}\frac{\triangle}{a^{2}}\dot{h}^{ij} +c_1 h_{ij}\frac{\triangle^2}{a^{2}} h^{ij}  \right).\nb\\
\end{eqnarray}
Here $\triangle \equiv \delta^{ij}\partial_{i}\partial_{j}$ with $\delta^{ij}$ being the Kronecker delta and a dot denotes the derivative respect to the cosmic time $t$.

Variation of the quadratic action (\ref{quadratic}) with respect to $h_{ij}$, one obtains the equation of motion for $h_{ij}$ as 
\begin{eqnarray} 
&&\left(1-c_1\frac{\partial^{2}}{a^{2}}\right)h^{\prime \prime}_{ij} + \left[2\mathcal{H}-c_1'\frac{\partial^{2}}{a^{2}}\right]h^{\prime}_{ij}\nonumber\\
&&-\left(1-c_1\frac{\partial^{2}}{a^{2}}\right)\partial^{2}h_{ij} = 0.
\end{eqnarray}
Then using the Fourier harmonics (\ref{Fourier}), the above equation can be cast into the form of eq.~(\ref{parametrized_eq}) as
\begin{eqnarray}
h^{\prime \prime}_{A} + (2 + {\bar \nu}) \mathcal{H} h^{\prime}_{A} + k^2 h_{A} = 0, \label{EoM_LV}
\end{eqnarray}
where
\bqn
{\cal H} \bar{\nu} = \left[\ln \left(1+c_1 \frac{k^2}{a^2}\right)\right]'.
\eqn
Considering that the effect from $c_1$ term is small, one approximately has
\bqn
{\cal H} \bar{\nu} \simeq \left(c_1 \frac{k^2}{a^2}\right)'.
\eqn
Then, one can connect the coupling coefficient $c_1$ in the action (\ref{action_ADM}) to the parameters $\alpha_{\bar \nu}$ and $M_{\rm LV}$ via
\begin{eqnarray}
   c_1(\tau) = \frac{\alpha_{\bar \nu}(\tau)}{M_{\rm LV}^2},
\end{eqnarray}
with $\beta_{\bar \nu}=2$. 
%Considering the coupling coefficient as a constant, which is equivalent to treating $\alpha_{\bar \nu}$ as constant, we can simply set $\alpha_{\bar \nu}=1$ by absorbing its order of magnitude into $M_{\rm LV}$.

In this paper, we consider the case with $\beta_{\bar \nu}=2$ and derive the corresponding modified waveform of GWs. We then explore the potential constraints on this modified waveform using proposed GW detectors such as aLIGO, CE, ET, Taiji, TianQin, and LISA. This case is induced by $\nabla_k K_{ij} \nabla^k K^{ij}$ which contains two time derivatives and two spatial derivatives. The cases with $\beta_{\bar \nu} > 2$ are also possible if one added terms with two time derivatives and more than two spatial derivatives in the gravitational action. However, these higher spatial derivative terms are expected to be suppressed, comparing to the leading-order case with $\beta_{\bar \nu}=2$. Therefore, in this paper, we only focus on the leading order one with $\beta_{\bar \nu}=2$.

\subsection{Amplitude modulation of GWs with Lorentz-violating damping effect}

The nonzero parameter $\overline{\nu}$ provides a frequency-dependent damping of GW amplitudes during propagation. This means that GWs with different frequencies will experience different damping rates. This damping rate effect induces an amplitude modification in the GWs. 

To study the modified waveform of GWs with this frequency-dependent damping of GW amplitudes, following the derivations in Refs. \cite{Zhao:2019xmm, Qiao:2019wsh}, let us decompose $h_A$ in Eq.~(\ref{EoM_LV}) as
\begin{eqnarray}
    h_A = h_A^{\rm GR} e^{-i\theta(\tau)}, \;\; h_A^{\rm GR}={\cal A}_A^{\rm GR} e^{-i \Phi^{\rm GR}(\tau)}.\label{decom}
\end{eqnarray}
Here $h_A^{\rm GR}$ is the solution of Eq. (\ref{EoM_LV}) when ${\bar \nu=0}$. ${\cal A}_A^{\rm GR}$ and $\Phi^{\rm GR}(\tau)$ are the amplitude and phase of $h_A^{\rm GR}$ respectively. With this decomposition, $\theta(\tau)$ encodes the correction arising from ${\bar \nu}$ which characterizes the Lorentz-violating damping effect. We would like to mention that, to obtain a waveform model with the propagation effects due to both the Lorentz-violating damping effect, we assume that the waveform extracted in the binary’s local wave zone is well-described by a waveform in GR. The same assumption has also been used in the analysis for testing the propagation effects in \cite{LIGOScientific:2019fpa, LIGOScientific:2020tif}. In this way, one can calculate both the amplitude and phase corrections due to the propagation effects to the GR-based waveform by using the stationary phase approximation (SPA) during the inspiral phase of the binary system \cite{Zhao:2019xmm}.

Plugging the second equation of decomposition Eq. (\ref{decom}) into Eq. (\ref{EoM_LV}) with ${\bar \nu=0}$, one finds
\begin{eqnarray}
    i \Phi'' + \Phi'^2+ 2 i {\cal H} \Phi'-k^2=0.
\end{eqnarray} 
Similarly, plugging the first equation of the decomposition Eq. (\ref{decom}) into Eq. (\ref{EoM_LV}), one gets
\begin{eqnarray}
    i (\theta'' + \Phi'')+ (\Phi'+\theta')^2 + i (2+{\bar \nu}) {\cal H}(\theta'+\Phi') - k^2=0.\nonumber \label{Phi_theta}\\
\end{eqnarray}
In GR, the time derivative of the phase $\Phi' \sim k$. Here the wavenumber $k$ is connected to the frequency of GWs by $k=2 \pi f/a_0$. Since the amplitude correction $\theta$ is induced by the expansion of the universe, one has $\theta' \sim {\cal H}$ and $\theta'' \sim {\cal H}^2$. Considering $k \gg {\cal H}$ and $\theta'' \ll \Phi' \theta' \sim k \theta'$, Eq. (\ref{Phi_theta}) can be simplified into 
\begin{eqnarray}
    2 \theta' + i {\cal H} {\bar \nu} \simeq 0.
\end{eqnarray}
Solving this equation gives
\begin{eqnarray}
    \theta = - \frac{i}{2} \int_{\tau_e}^{\tau_0} {\cal H} {\bar \nu} d\tau.
\end{eqnarray}
The Lorentz-violating damping effect in the phase $\theta$ is purely imaginary, indicating that it modifies the amplitude of the GWs during their propagation. Considering ${\bar \nu}$ is also frequency-dependent, such amplitude modulation depends on the frequency of GWs as well. 

Specifically, with the above solution, one can write the waveform of GWs with Lorentz-violating effect as
\begin{eqnarray}
    h_{A} = h_{A}^{\rm GR} e^{\delta h_2}, \label{waveform_0}
\end{eqnarray}
where
\begin{eqnarray}
    \delta h_2 &=& - \frac{1}{2} \int_{\tau_e}^{\tau_0} {\cal H}\bar \nu d\tau \nonumber \\
    &=& - \frac{1}{2} \left[\alpha_{\bar \nu} \left(\frac{k}{a M_{\rm LV}}\right)^{\beta_{\bar \nu}}\right]\Bigg|^{a_0}_{a_e}.
\end{eqnarray}
It can be further rewritten in the form
\begin{eqnarray}
    \delta h_2 = - \frac{1}{2} \left(\frac{2 u}{M_{\rm LV} {\cal M}}\right)^{\beta_{\bar \nu}} \Big[\alpha_{\bar \nu}(\tau_0) - \alpha_{\bar \nu}(\tau_e) (1+z)^{\beta_{\bar \nu}}\Big].\nonumber \label{deltah2}\\
\end{eqnarray}
Here we define $u= \pi {\cal M} f$, where ${\cal M} = (1+z) {\cal M}_{\rm c}$ represents the measured chirp mass, and  ${\cal M}_{\rm c} \equiv (m_1 m_2)^{3/5}/(m_1+m_2)^{1/5}$ denotes the chirp mass of the binary system with component masses $m_1$ and $m_2$.

\subsection{Amplitude modification to the waveform of GWs}

To derive the modified waveform of GWs, we consider the GWs produced during the inspiral stage of the compact binaries. To directly contact with the observations, it is convenient to analyze the GWs in the Fourier domain. In this approach, under the stationary phase approximation, the responses of the detectors to the GW signal ${\tilde h}(f)$ can be written in the form of
\begin{eqnarray}
    {\tilde h}(f) = \Big[F_+ h_+(f) + F_{\times} h_{\times}(f)\Big] e^{-2 \pi i f \Delta t},
\end{eqnarray}
where $F_{+}$ and $F_{\times}$ denote the beam pattern functions of GW detectors, which depend on the GW source's location and polarization angle \cite{Jaranowski:1998qm, Zhao:2017cbb}. The two polarizations of GWs, $h_{+}(f)$ and $h_{\times}(f)$, are related to the left- and right-handed polarization modes, $h_{\rm L}(f)$ and $h_{\rm R}(f)$, via
\begin{eqnarray}
    h_+ = \frac{h_{\rm L}+h_{\rm R}}{\sqrt{2}}, \;\; h_{\times} = \frac{h_{\rm L} - h_{\rm R}}{\sqrt{2}i}.
\end{eqnarray}
Then using Eq. (\ref{waveform_0}), after tedious calculations, one obtains the following restricted form for the waveform of GWs in the Fourier domain as a function of the GW frequency $f$, i.e., 
\begin{eqnarray}
     \tilde{h}(f)=\mathcal{A}_{\rm GR}f^{-7/6}e^{i\Psi_{\mathrm{GR}}(f)}e^{\delta h_{2}}, \label{waveforms}
\end{eqnarray}
where ${\cal A}_{\rm GR}$ and $\Psi_{\rm GR}$ represent the amplitude and phase of GWs of a compact binary inspiral signal in GR, and $e^{\delta h_2}$ with $\delta h_2$ being given by Eq. (\ref{deltah2}) is the amplitude correction to the waveform of GWs in GR. In the post-Newtonian approximation, the amplitude and the phase of GWs in GR can be expressed as \cite{Wang:2022yxb}
\begin{eqnarray}
    {\cal A}_{\rm GR} = \frac{2}{5} \times \sqrt{\frac{5}{24}} \pi^{-2/3} \frac{\mathcal{M}^{5/6}}{D_{\rm L}},
\end{eqnarray}
and
\begin{widetext}
\begin{eqnarray}
    \Psi_{\mathrm{GR}}(f) &=& 2 \pi f t_{\rm c}-\Phi_{\rm c}-\frac{\pi}{4}+\frac{3}{128 \eta} u^{-5 / 3}\left\{1+\left(\frac{3715}{756}+\frac{55}{9} \eta\right) u^{2 / 3}-16 \pi u\right.\nonumber \\
    && +\left(\frac{15293365}{508032}+\frac{27145}{504} \eta+\frac{3085}{72} \eta^2\right) u^{4 / 3}+\pi\left(\frac{38645}{756}-\frac{65}{9} \eta\right)(1+\ln u) u^{5 / 3}\nonumber \\
    && +\left[\frac{11583231236531}{4694215680}-\frac{640}{3} \pi^2-\frac{6848}{21} \gamma_{\mathrm{E}}-\frac{6848}{63} \ln (64 u)+\left(-\frac{15737765635}{3048192}+\frac{2255}{12} \pi^2\right) \eta\right.\nonumber \\
    && \left.\left.+\frac{76055}{1728} \eta^2-\frac{127825}{1296} \eta^3\right] u^2+\pi\left(\frac{77096675}{254016}+\frac{378515}{1512} \eta-\frac{74045}{756} \eta^2\right) u^{7 / 3}\right\},
\end{eqnarray}
\end{widetext}
where the luminosity distance $D_{\rm L}$ is expressed as
\begin{eqnarray}
    D_{\rm L}(z)=\frac{1+z}{H_{0}} \int_{0}^{z} \frac{dz^{\prime}}{\sqrt{\Omega_{\mathrm{M}}(1+z^{\prime})^{3} +\Omega_{\Lambda}}}.
\end{eqnarray}
Here we adopt the $\Lambda$CDM model with Hubble constant $H_{0} \approx67.4$ km $\mathrm{s}^{-1}$ $\mathrm{Mpc}^{-1}$, matter density fractoin $\Omega_{\mathrm{m}} \approx 0.315$, and vacuum energy density fraction $\Omega_{\mathrm{\Lambda}} \approx 0.685$ \cite{Planck:2018vyg, Zhang:2019ylr}. $t_c$ and $\Phi_c$ are time and phase at coalescence, $\gamma_{\rm E}$ is the Euler constant, and $\eta=m_1 m_2/(m_1+m_2)^2$ is the symmetric mass ratio.

\begin{table}
\caption{\label{tab:table1}%
Characteristics of six GW detectors.}
\begin{ruledtabular}
\begin{tabular}{cccccc}
Detector & Configuration & $f_{\rm lower}$ [Hz]& $f_{\rm upper}$ [Hz]& Reference\\
\hline
aLIGO & Rightangle & 10 & 5000 & Refs. \cite{Wang:2022yxb,KAGRA:2023pio}\\
ET & Rightangle & 1 & 10000 & Ref. \cite{Hild:2010id} \\
CE & Rightangle & 5 & 4000 & Ref. \cite{Reitze:2019iox}\\
Taiji & Triangle & 0.0001 & 0.1 & Ref. \cite{Liu:2023qap} \\
TianQin & Triangle & 0.0001 & 1 & Ref. \cite{TianQin:2020hid} \\
LISA & Triangle & 0.0001 & 0.1 & Refs. \cite{Liu:2023qap,LISA:2017pwj} \\
\end{tabular}
\end{ruledtabular}
\end{table}

\section{ANALYSIS FRAMEWORK WITH FISHER INFORMATION MATRIX}\label{sec:3}
\renewcommand{\theequation}{3.\arabic{equation}} \setcounter{equation}{0}

\subsection{General considerations}

In this section, we provide a brief overview of the matched-filter analysis with the FIM approach, which follows the method outlined for compact binary inspiral in Refs. \cite{Cutler:1994ys, Finn:1992xs, Poisson:1995ef}. We calculate the noise-weighted inner product between the partial derivatives of each GW waveform parameter and the one-sided power spectral density (PSD) of the detector noises. This calculation yields the FIM. Inverting the FIM provides the variance-covariance matrix, where the diagonal elements represent the square root of the mean squared error for the estimated parameters of the signal. Previous studies have showcased the precision and utility of the FIM approach, particularly in situations with high signal-to-noise ratios (SNRs).

To be specific, we first give the noise-weighted inner product of two signals $h_{1}$ and $h_{2}$ 
\begin{equation}
    \left({h}_{1}|{h}_{2}\right)= 2 \int_{f_{\mathrm{min}}}^{f_{\mathrm{max}}}\frac{\tilde{h}_{1}^{*}(f)\tilde{h}_{2}(f)+\tilde{h}_{2}^{*}(f)\tilde{h}_{1}(f)}{S_{\rm n}(f)} df,
\label{inner product}
\end{equation}
where $\tilde{h}_1(f)$ and ${\tilde h}_2(f)$ are the Fourier transformation of GW signal $h(t)$, $S_{\rm n}(f)$ is the PSD of the detector's noise, and the star superscript stands for complex conjugation. In the above expression, $f_{\rm max}\ (f_{\rm min})$ represents the instrumental maximum (minimum) threshold frequency.

For a given signal $h$, the SNR is defined as
\begin{equation}
    \rho=(h|h)^{1/2}.
\end{equation}
The modified waveform of GW from binary inspirals, influenced by the Lorentz-violating damping effect as described in Eq. (\ref{waveforms}), are generally characterized by a set of parameters $\theta_i$. In this context, one can define the FIM as
\begin{equation}
F_{ij}=\left(\left.\frac{\partial \tilde{h}}{\partial \theta_{i}}\right|\frac{\partial \tilde{h}}{\partial \theta_{j}}\right).
\end{equation}
Here $\theta_{i}$ and $\theta_{j}$ represent the elements in the set of modified waveform parameters of GWs $\{\ln \mathcal{A}, \ln \mathcal{M}, \ln \mathcal{\eta}, \phi_{\mathrm{c}}, t_{\mathrm{c}}, C_{\nu} \}$, where $C_{\nu}=M_{\mathrm{LV}}^{-2}$ characterizes the Lorentz-violating damping effect in the waveform. Then one can calculate each element of the FIM, which are given respectively in Appendix A.

In the large SNR approximation, if the noise is stationary and Gaussian, the probability that the GW signal $h(t)$ can be characterized by a given set of values of the parameters $\theta_i$, 
\bqn
p(\theta_i|h)=p^{(0)}(\theta_i) \exp\left[-\frac{1}{2}F_{ij} \Delta \theta^i \Delta \theta^j\right],
\eqn
where $p^{(0)}(\theta_i)$ represents the distribution of prior information. Then, the standard deviations $\Delta\theta_i$ in measuring the parameter $\theta_i$, which mean $1\sigma$ bounds on parameters, can be calculated in the large SNR approximation. This can be obtained by taking the square root of the corresponding diagonal elements in the inverse of FIM,
\begin{equation}
    \Delta \theta_{i} =\sqrt{(F^{-1})_{ii}},
\end{equation}
where $F^{-1}$ is the inverse of FIM.

Our main purpose in employing FIM analysis here is to gauge the potential of future GW detectors in constraining the energy scale $M_{\rm{LV}}$ associated with Lorentz violation, which induces the Lorentz-violating damping effect in the propagation of GWs. We consider two types of GW detectors, ground-based GW detectors including aLIGO, CE, and ET,  and space-based GW detectors, including Taiji, TianQin, and LISA. We summarize the information of all the six GW detectors in Table \ref{tab:table1}.

To investigate the variation tendency of $M_{\rm LV}$ in GW events, we choose the reasonable astrophysical horizon
of each GW detector as the sources' property and constrain $M_{\rm LV}$ by FIM. To maintain the effectiveness of FIM, the ranges of redshift $z$ and total mass $M$ are suitable enough to satisfy the SNR threshold ($\rho > 8$ for ground-based detectors and $\rho > 15$ for space-based detectors). For ground-based GW detectors, their detectable frequency bands are several Hz to thousand Hz. As the third generation GW detector, the frequency band of CE can reach $[5-4000]$ Hz. For space-based detectors, with huge arm-lengths so that they can detect the GWs from compact binary systems in a very low-frequency band $[10^{-4}-10^{-1}]$ Hz. The compact binary systems in this low-frequency band usually consist of supermassive black holes and their total mass range is $[10^{4}-10^{6}]\ \mathrm{M}_{\odot} $.

When calculating the integrals of the inner product, it is necessary to use the appropriate limits of integration. The minimum frequency $f_{\mathrm{min}}$ in Eq. (\ref{inner product}) is taken as the instrumental minimum threshold frequency of the GW detector as shown in Table \ref{tab:table1}. The upper cut-off frequency $f_{\mathrm{max}}$ is chosen from $\mathrm{min}\{f_{\mathrm{upper}}, f_{\mathrm{ISCO}}\}$, where $f_{\mathrm{upper}}$ is the upper frequency of detectors and $f_{\mathrm{ISCO}}$ is usually estimated by the innermost stable circular orbit (ISCO) \cite{Mirshekari:2011yq}
\begin{eqnarray}
    f_{\mathrm{ISCO}} = 6^{-3/2} \pi^{-1} \eta^{3/5} \mathcal{M}^{-1}.
\label{fisco}
\end{eqnarray}
When calculating the FIM, we set $C_{\nu} = 0$ as the fiducial value. Additionally, the values of $t_{\rm c}$ and $\phi_{\rm c}$ do not impact the constraints on other parameters; hence, we set $t_{\rm c} = 0$ and $\phi_{\rm c} = 0$.

\subsection{Noise power spectral density of detector}

In Fig. \ref{fig:detectors}, we illustrate the noise PSD of both ground-based and space-based detectors. The three ground-based detectors' PSD are from the official data files quoted in Table \ref{tab:table1}. The other three PSD data are from the theoretical formula introduced below.

The sensitivity curve of Taiji can be obtained by the formula \cite{Liu:2023qap}
\begin{align}
    S_{\rm n}(f) &= \frac{10}{3 L^{2}} \left(P_{\rm dp} + 2 (1+\mathrm{cos}^{2}(f/f_{*})) \frac{P_{\rm acc}}{(2 \pi f)^{4}}\right)  \nonumber\\
    &\times \left(1+0.6 (\frac{f}{f_{*}})^{2}\right),
\end{align}
where 
\begin{equation}
P_{\mathrm{dp}}=\left(8 \times 10^{-12} \mathrm{~m}\right)^2\left(1+\left(\frac{2\, \mathrm{mHz}}{f}\right)^4\right) \mathrm{Hz}^{-1},
\end{equation}
\bqn
P_{\mathrm{acc}}&=&\left(3 \times 10^{-15} \mathrm{~m} \mathrm{~s}^{-2}\right)^2\left(1+\left(\frac{0.4\, \mathrm{mHz}}{f}\right)^2\right) \nonumber\\
&&\times\left(1+\left(\frac{f}{8\, \mathrm{mHz}}\right)^4\right) \mathrm{Hz}^{-1}.
\eqn
Here $P_{\rm dp}$ is the PSD of the displacement noise and $P_{\rm acc}$ is the PSD of the acceleration noise. $f_{*} = 1/(2 \pi L)$ and $L$ is the arm-length of the detector. For Taiji $L = 3 \times 10^{9}\ \rm m$.

For LISA, $L = 2.5 \times 10^{9}$ m and the displacement noise can be written as \cite{Liu:2023qap, Berti:2004bd}
\begin{equation}
P_{\mathrm{dpL}}=\left(15 \times 10^{-12} \mathrm{~m}\right)^2\left(1+\left(\frac{2\, \mathrm{mHz}}{f}\right)^4\right) \mathrm{Hz}^{-1}.
\end{equation}

The sensitivity curve for TianQin can be modeled by the following equation \cite{TianQin:2020hid}
\bqn
S_{\rm n}(f)&=&\frac{10}{3 L^2}\left[S_{\rm x}+\frac{4 S_{\rm a}}{(2 \pi f)^4}\left(1+\frac{10^{-4} \mathrm{~Hz}}{f}\right)\right]\nonumber\\
&&\times\left[1+0.6\left(\frac{f}{f_*}\right)^2\right],
\eqn
where displacement measurement noise $S^{1/2}_{\rm x}$ and residual acceleration noise $S^{1/2}_{\rm a}$ are defined as
\begin{eqnarray}
    S^{1/2}_{\rm x} = 1\times10^{-12}\ \mathrm{m} /\mathrm{Hz}^{1/2},
\end{eqnarray}
and
\begin{eqnarray}
    S^{1/2}_{\rm a} = 1\times10^{-15}\ \mathrm{m} \ \mathrm{s}^{-2}/\mathrm{Hz}^{1/2}.
\end{eqnarray}
$f_{*} = c/(2 \pi L)$ is the transfer frequency and $L = \sqrt{3}\times 10^{8}\ \mathrm{m}$.

\section{RESULTS AND DISCUSSION}\label{sec:4}
\renewcommand{\theequation}{4.\arabic{equation}} \setcounter{equation}{0}

\begin{table*}
\caption{\label{tab:table2}The best constraints of the simulated single GW events and the combination of joint events from each GW detector. The results from the single event are the best constraint which are from Fig. \ref{fig:mlv for six detectors}. The redshift ranges and the total mass ranges are only for the joint events. The number of joint events contains 100 (for both ground-based and space-based detectors) simulated GW events. We choose $\rho > 8$ and $\rho > 15$ as the threshold for ground-based detectors and space-based detectors respectively.}
\begin{ruledtabular}
\begin{tabular}{cccccc}
Detector&Single event [Gev]
&Joint redshift range&Joint total mass range [$M_{\odot}$]&Joint number&Joint event [Gev]\\ \hline
aLIGO&$8.54 \times 10^{-22}$&[0.01-0.5]&[3-10]& 100 &$2.39 \times 10^{-21}$\\
CE&$3.02 \times 10^{-21}$&[0.01-0.5]&[3-10]& 100 &$8.84 \times 10^{-21}$\\
ET&$2.39 \times 10^{-21}$&[0.01-0.5]&[3-10]& 100 &$7.18 \times 10^{-21}$\\
Taiji&$2.10 \times 10^{-24}$&[5-10]&$[4.2-9.2]\times 10^{3}$& 100 &$5.86 \times 10^{-24}$\\
TianQin&$1.89\times 10^{-24}$&[0.01-5]&$[1.5-2]\times 10^{4}$& 100 &$4.56 \times 10^{-24}$\\
LISA&$1.52\times 10^{-24}$&[5-10]&$[8.3-13.3]\times 10^{3}$& 100 &$4.37 \times 10^{-24}$\\
\end{tabular}
\end{ruledtabular}
\end{table*}

\begin{figure*}
\centering
\includegraphics[width=8.3cm]{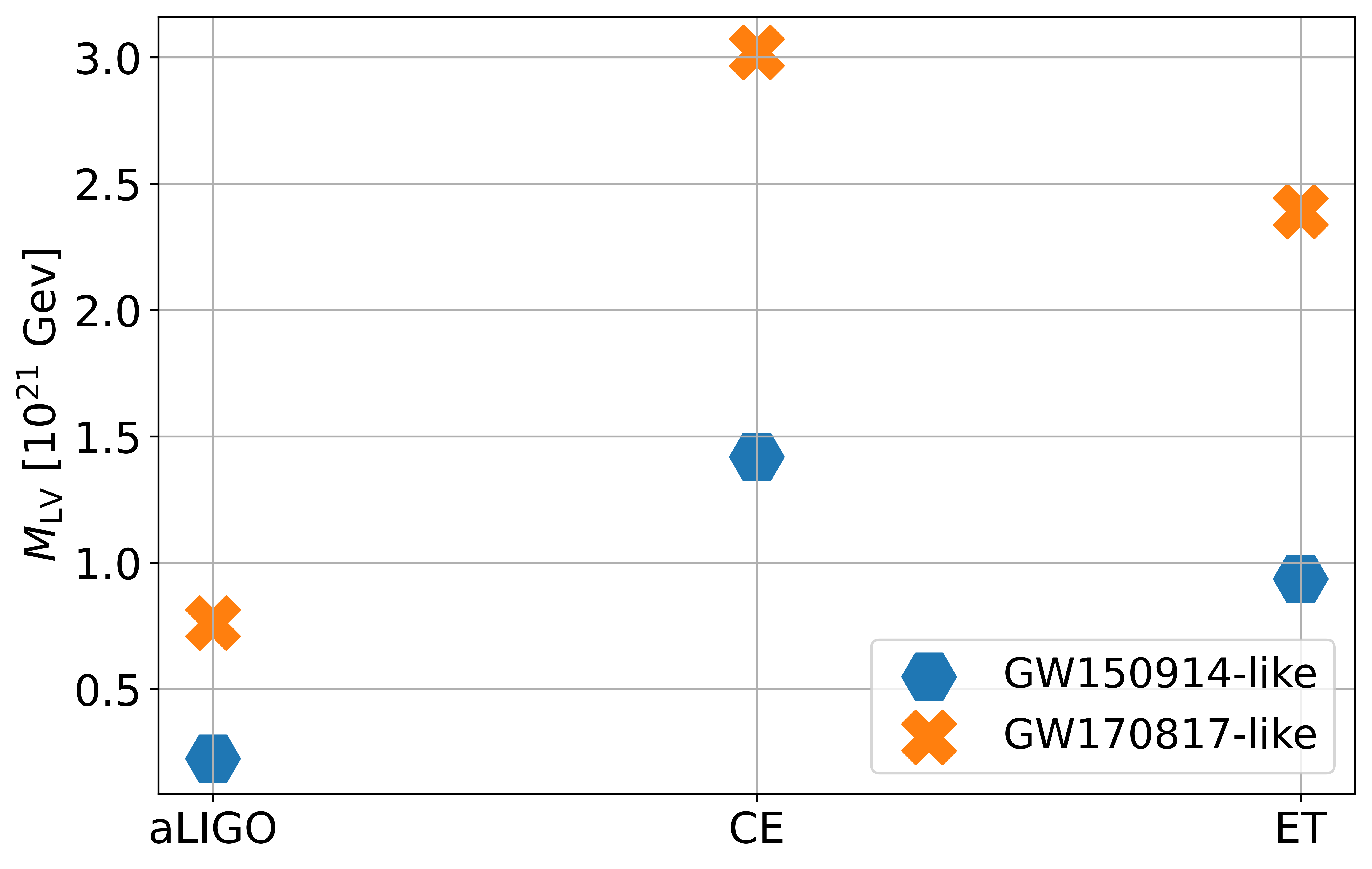}
\includegraphics[width=8.1cm]{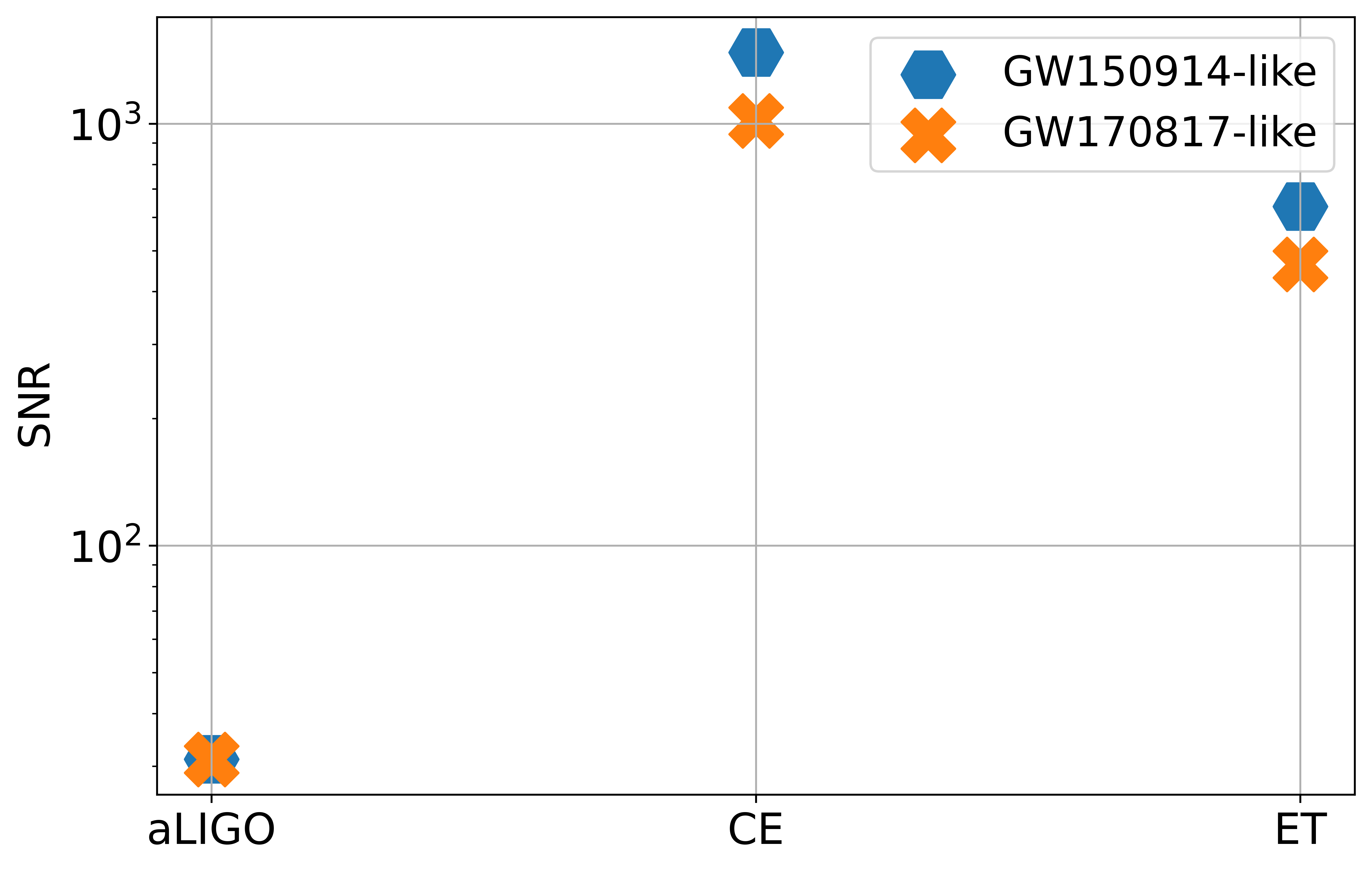}
\caption{The results of $M_{\rm LV}$ and SNR of GW150914-like and GW170817-like events in three ground-based detectors. GW170817-like event is the compact binary system of two neutron stars and GW150914-like event is a system containing two black holes.}
\label{fig:GW_events.png}
\end{figure*}

\begin{figure*}
\centering
\includegraphics[width=5.8cm]{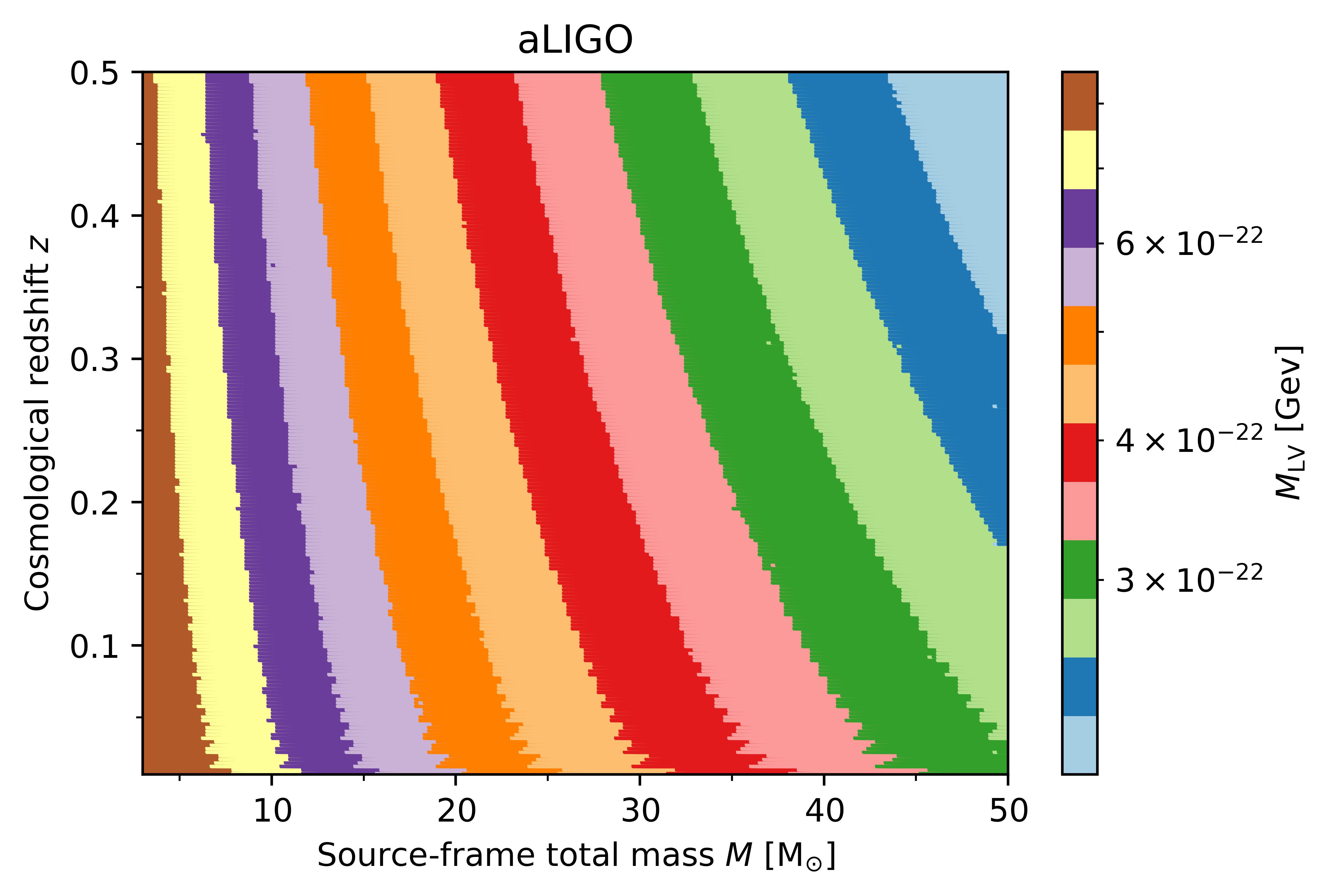}
\includegraphics[width=5.8cm]{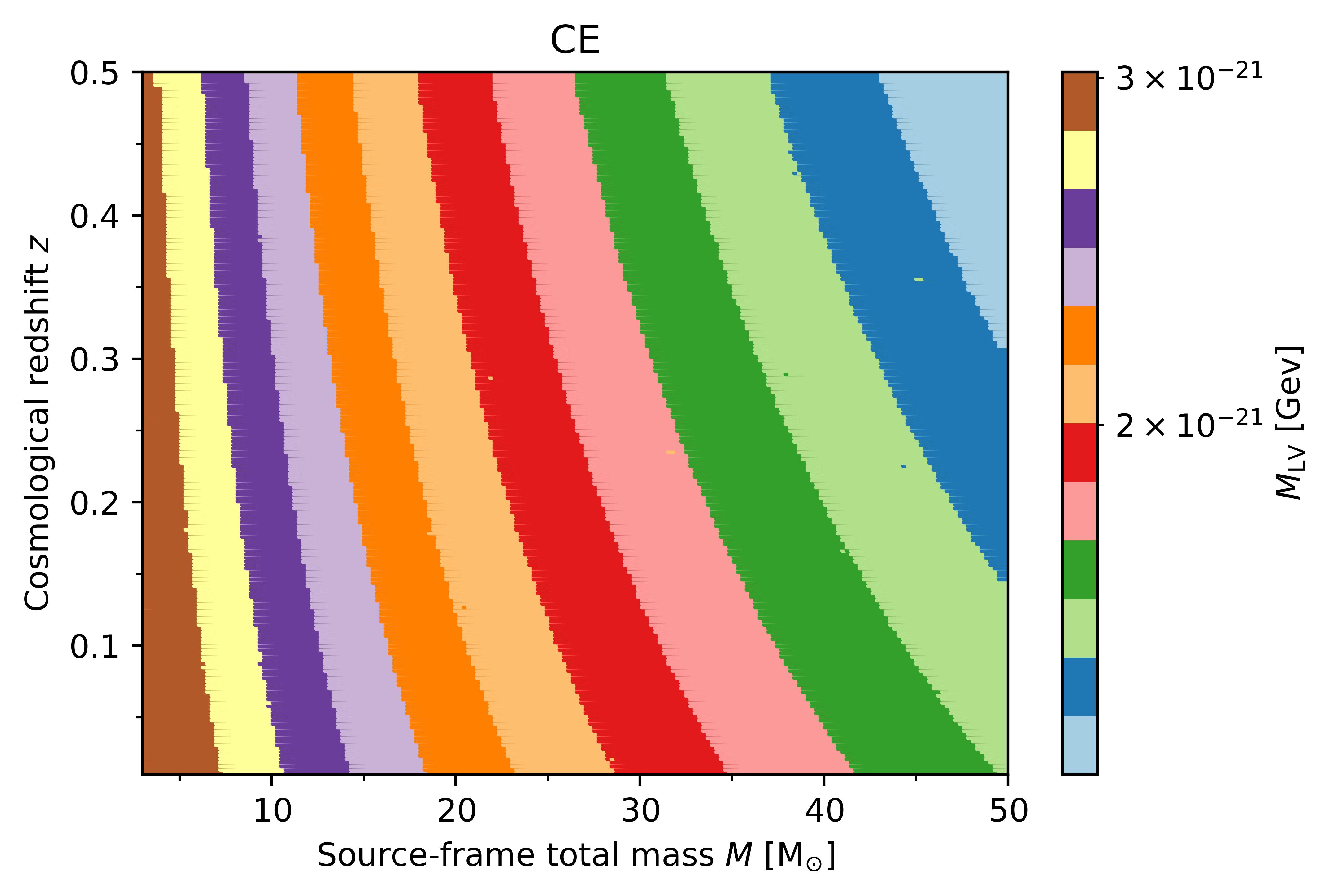}
\includegraphics[width=5.8cm]{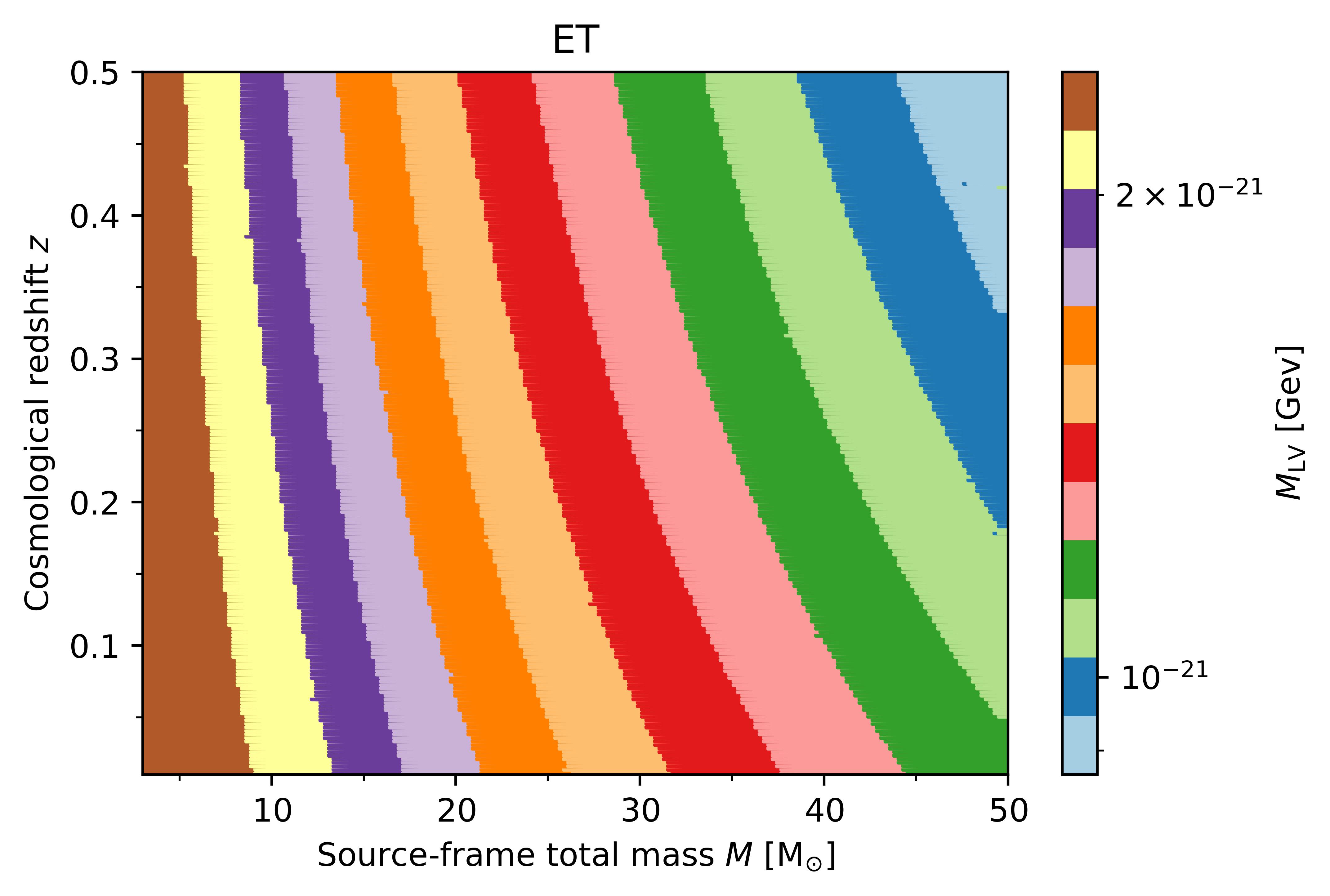}
\includegraphics[width=5.8cm]{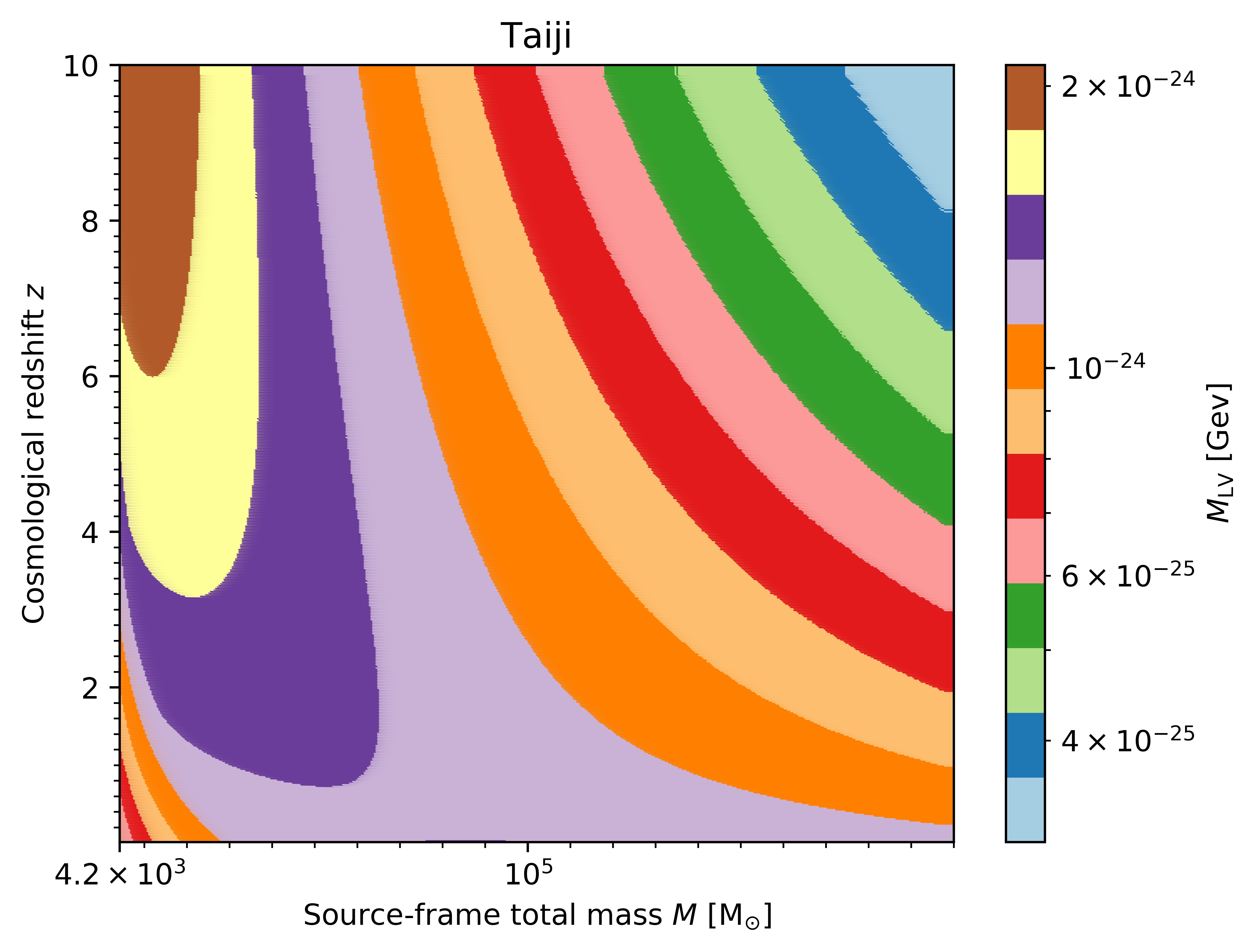}
\includegraphics[width=5.6cm]{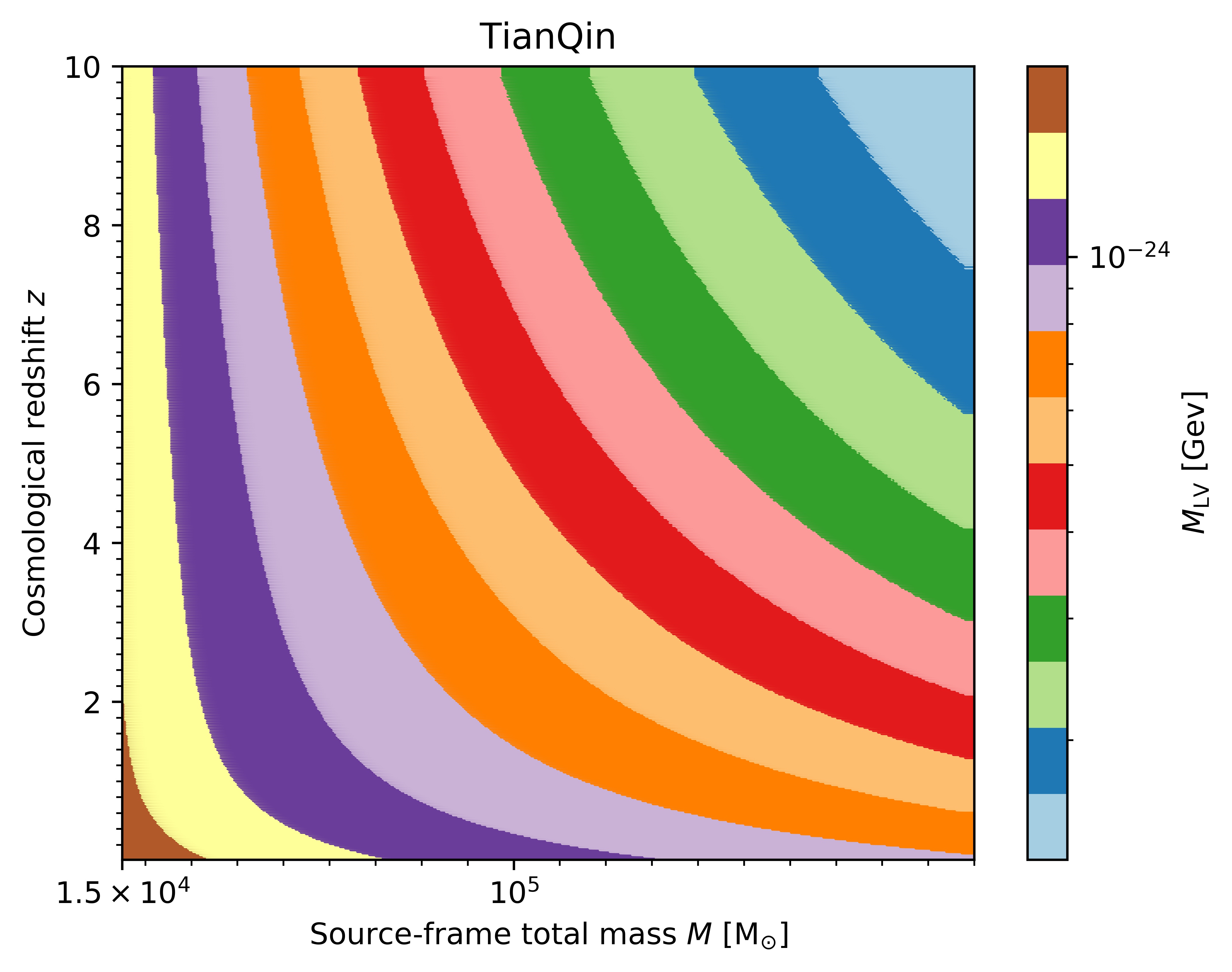}
\includegraphics[width=5.8cm]{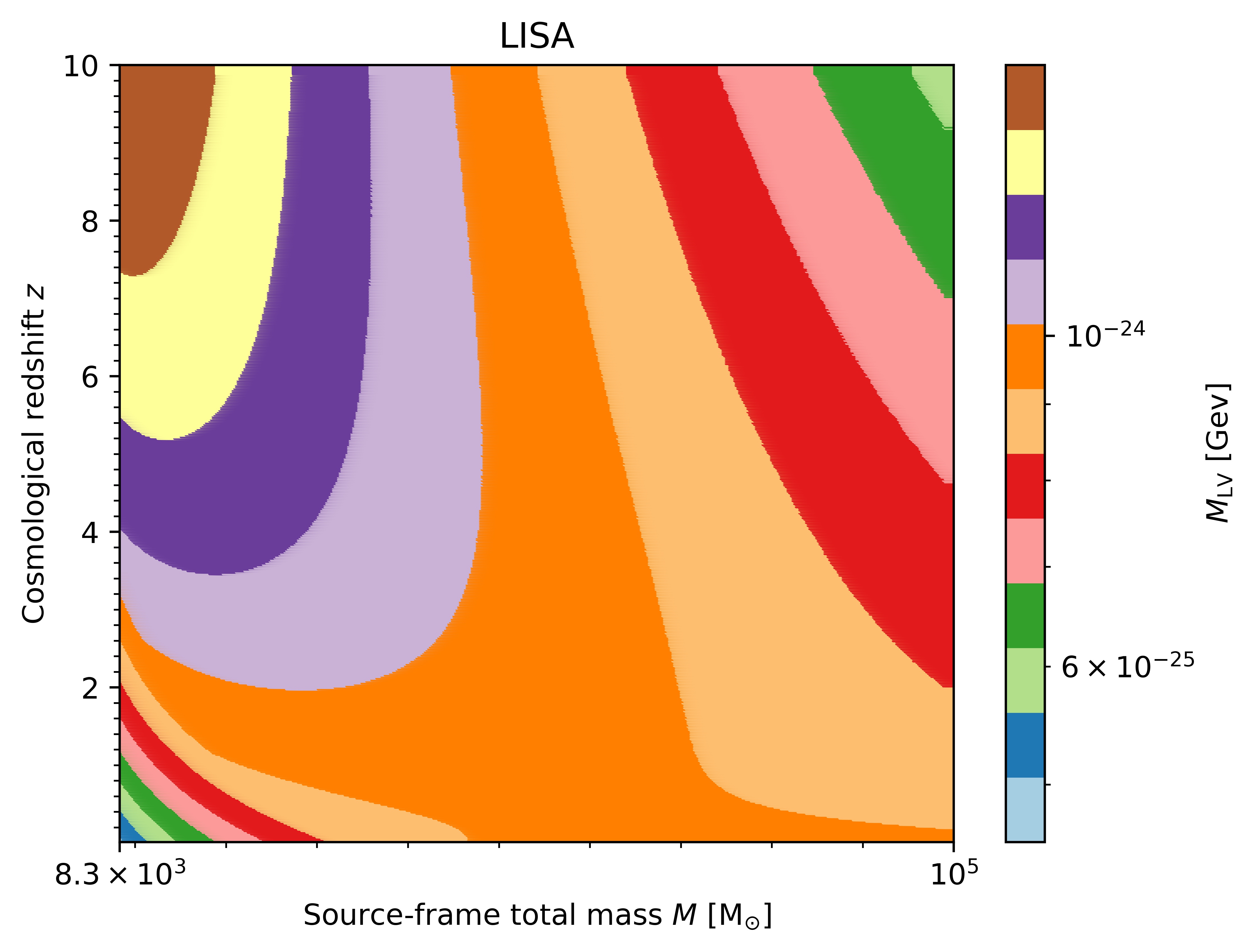}
\caption{Dependence of the lower bound of $M_{\rm LV}$ on the total mass and redshift of the binary inspiral systems for different detectors, aLIGO (top left), CE (top middle), ET (top and right), Taiji (bottom left), TianQin (bottom middle), and LISA (bottom right).
\label{fig:mlv for six detectors}}
\end{figure*}

In this section, we present the results of the potential constraints on the parameters of the modified waveform of GWs with the Lorentz-violating damping effect. Among these parameters, we focus on the energy scale $M_{\rm LV}$ of Lorentz violation which characterizes the frequency-dependent damping of GWs during their propagations. To conduct a comprehensive and reliable analysis of GW parameter estimation, we consider the frequency bands of both ground-based and space-based detectors. We employ simulated GW data to constrain the Lorentz-violating damping effect through the FIM approach. For the simulated GW data, we refer to the astrophysical horizons of the detectors from Ref. \cite{Evans:2021gyd} (for aLIGO, ET), Refs. \cite{Srivastava:2022slt, Evans:2021gyd} (for CE), Ref. \cite{Liu:2023qap} (for Taiji), Ref. \cite{TianQin:2020hid} (for TianQin), and Ref. \cite{LISA:2017pwj} (for LISA). We set $\rho > 8$ and $\rho > 15$ as the thresholds for ground-based detectors and space-based detectors, respectively. To fully exploit the high event rates of future GW detectors, we also decide to conduct a multi-event joint constraint analysis. The results of constraining both individual GW events and their combinations are detailed in Table \ref{tab:table2}. The constraint results and SNR from GW150914-like and GW170817-like events are shown in Fig. \ref{fig:GW_events.png}. We depict the dependence of the lower bound of $M_{\mathrm{LV}}$ in Fig. \ref{fig:mlv for six detectors}.

\subsection{$M_{\rm LV}$ from ground-based detectors}

First, let us analyze the constraints on $M_{\rm LV}$ from the three ground-based detectors. In Fig.~\ref{fig:GW_events.png}, we illustrate the constraints on $M_{\rm LV}$ (right panel) and SNRs (left panel) of two examples of GW events, the GW150914-like and GW170817-like events with three different ground-based detectors. As observed in the left panel of Fig.~\ref{fig:GW_events.png}, the lower bounds on $M_{\rm LV}$ from both CE and ET extend to $\gtrsim 10^{-21}$ GeV. CE also gives the best constraints on $M_{\rm LV}$ for both GW150914-like and GW170817-like events. Notably, under similar redshift conditions, the GW170817-like event, characterized by a smaller total mass, attains a higher lower bound of $M_{\rm LV}$. Moving to the right panel of Fig.~\ref{fig:GW_events.png}, CE exhibits a distinct advantage in SNR, suggesting its effectiveness in ensuring the detection of such GW events. There appears to be an indicative trend implying that GW events with a greater total mass could potentially yield higher SNRs.

In Table \ref{tab:table2}, it is evident that ground-based detectors, capable of observing high-frequency GWs, exhibit superior performance compared to their space-based counterparts. Among ground-based detectors, the strongest constraint for $M_{\rm LV}$ is achieved by the third-generation detector CE at $3.02 \times 10^{-21}$ GeV. The result from the ET is marginally smaller than that of CE. For aLIGO, the constraint can reach $8.54 \times 10^{-22}$ Gev. 

Considering the prospect of observing a large number of GW events in the future, we also do research on the joint analysis for individual detectors. As anticipated, the outcomes from each detector exhibit a marked enhancement. The most notable improvement comes from CE, yielding the optimal result with $M_{\mathrm{LV}} > 8.84 \times 10^{-21}$ GeV. We combine 100 simulated GW events within the redshift range [0.01-0.5] Hz and total mass range [3-10] $M_{\odot}$ of sources. aLIGO attains a result of $M_{\mathrm{LV}} > 2.39 \times 10^{-21} \mathrm{GeV}$. This value is approximately twice as high as the result reported in Ref. \cite{Zhu:2023wci}, which means that aLIGO continues to be a powerful tool in testing the Lorentz-violating damping effect. It is crucial to highlight that the arm-length of ET is 10 km. Meanwhile, CE is designed with arm-lengths of 20 km and 40 km. Due to their long arm-lengths and enhanced sensitivity, CE and ET can provide values of $M_{\mathrm{LV}}$ that are roughly more than double the value obtained by aLIGO. The expected detection rate, as reported in Refs. \cite{Li:2023gtu, Han:2023exn}, is conservatively estimated at 100.

\subsection{$M_{\rm LV}$ from space-based detectors}

For the space-based detectors, including Taiji, TianQin, and LISA, the constraints on $M_{\rm LV}$ are about three orders of magnitude weaker than those from the ground-based detectors. As shown in Table~\ref{tab:table2}, the constraints on $M_{\rm LV}$ is roughly at $\gtrsim 2 \times 10^{-24}\ {\rm GeV}$ for single GW event. When considering a joint analysis of 100 simulated GW events, the constraints from all three detectors are roughly at $\gtrsim 5 \times 10^{-24}\; {\rm GeV}$. Among the three detectors, Taiji achieves the best result, with $M_{\mathrm{LV}} > 5.86 \times 10^{-24}$ GeV. The reason why the constraints from the space-based detectors are weaker than those from the ground-based detectors is easy to understand. As one can see from Eq. (\ref{deltah2}), the amplitude correction to the waveform due to the Lorentz-violating damping effect is proportional to the square of the GW frequencies, which implies this effect is more sensitive to the higher GW frequencies, and thus the ground-based detectors can give stronger constraints than space-based detectors. We note that to estimate the number of events within our joint redshift and total mass range, we employ the data and methods from Refs. \cite{Wang:2021srv, Klein:2015hvg}.

\subsection{Trends of $M_{\rm LV}$}

\begin{figure}
%\centering
\includegraphics[width=8.3cm]{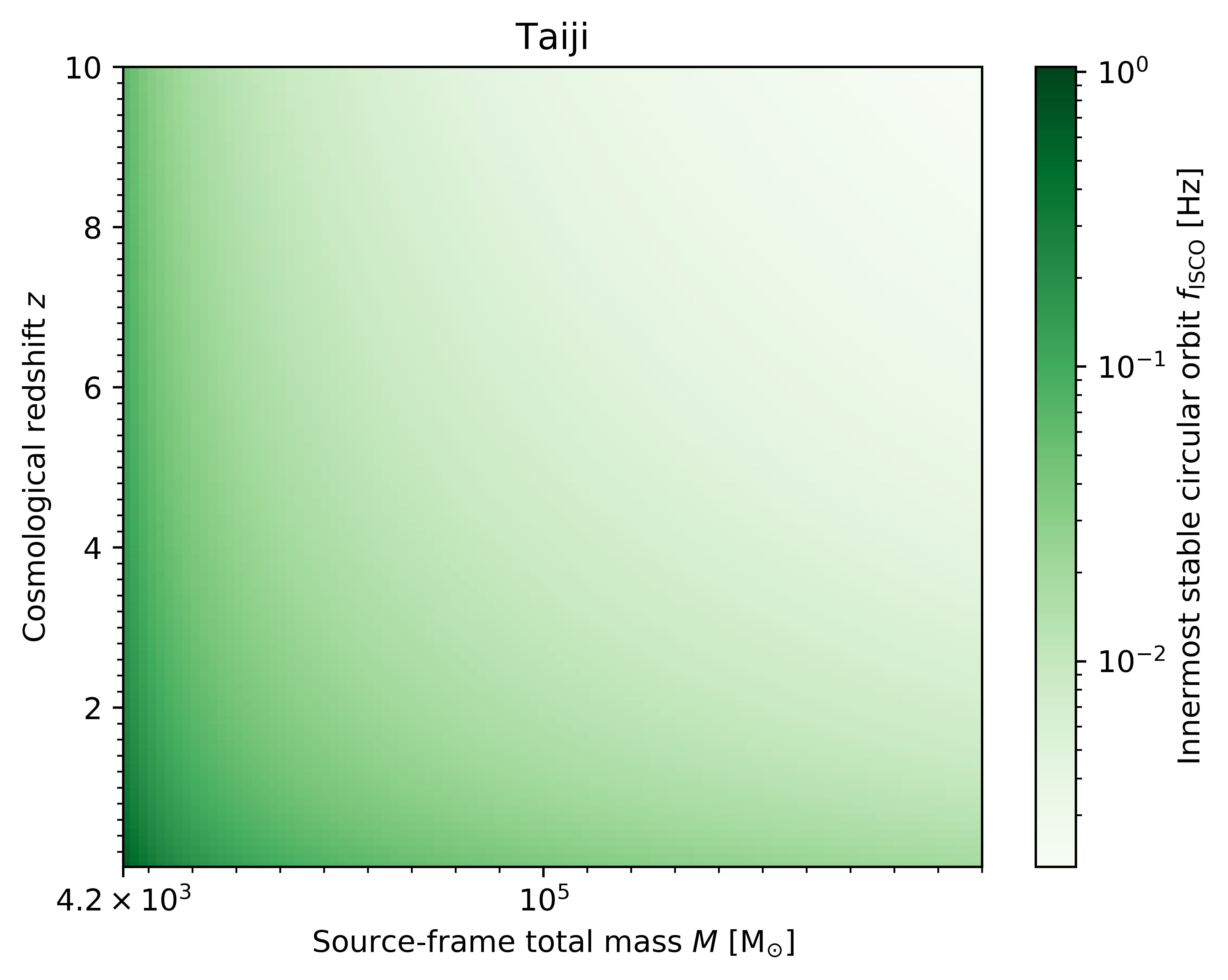}
\includegraphics[width=8.3cm]{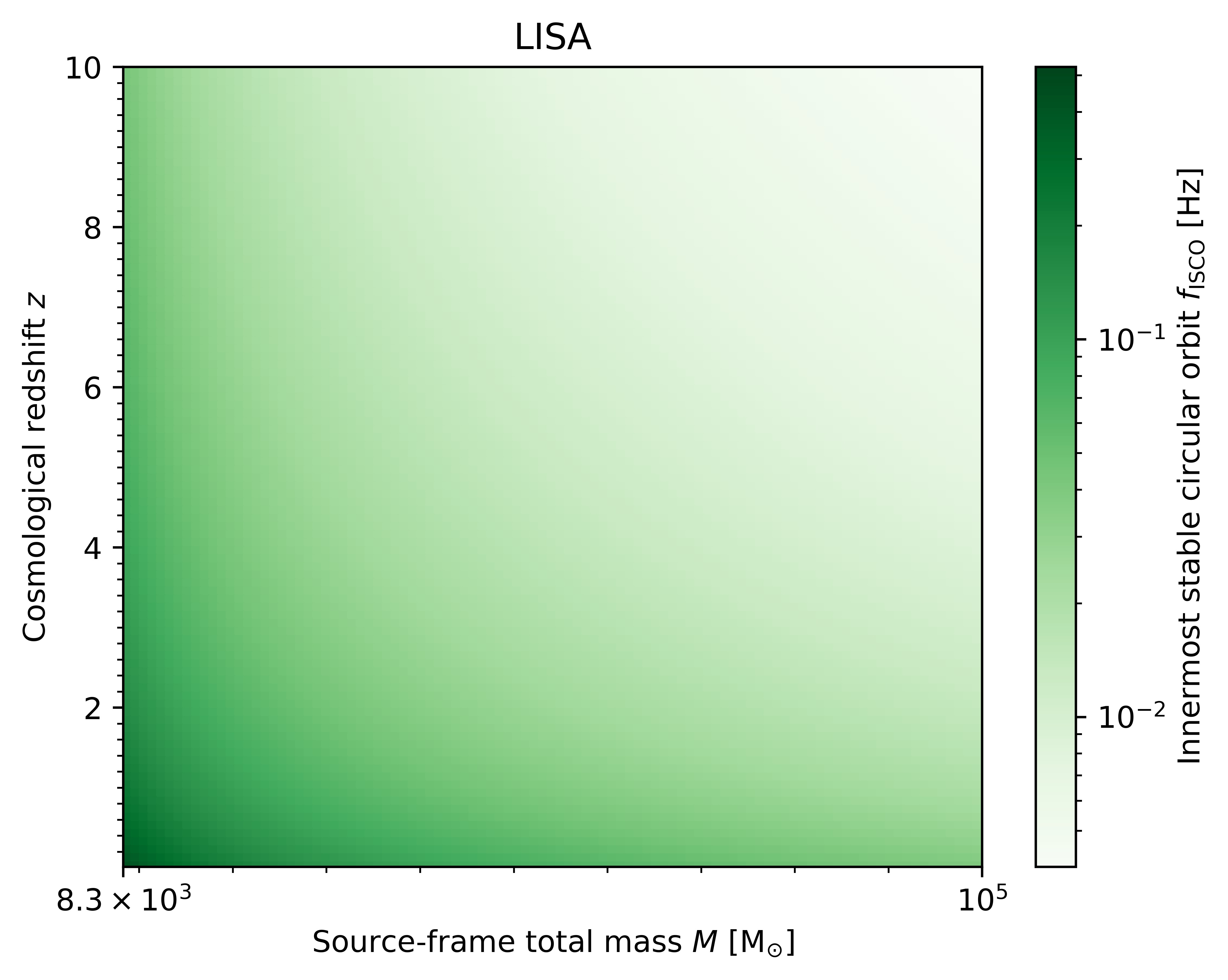}
\caption{The distribution of $f_{\mathrm{ISCO}}$ for Taiji and LISA is depicted within the same astrophysical horizon, as illustrated in Fig. \ref{fig:mlv for six detectors}.
\label{fig:fisco for Taiji and LISA}}
\end{figure}

In Fig. \ref{fig:mlv for six detectors}, we illustrate how the lower bound of the Lorentz-violating parameter, $M_{\rm LV}$, varies with the total mass and redshift of binary inspiral systems across a selection of detectors. Specifically, for aLIGO, CE, and ET, we focus on a redshift range of $[0.01, 0.5]$ Hz and a total mass range of $[3, 50]\ M_{\odot}$. For Taiji, TianQin, and LISA, we extend the redshift range to $[0.01, 10]$ Hz. The rationale behind selecting different total mass ranges for these latter three detectors is linked to ensuring that the SNR, $\rho$, exceeds 15 when the redshift value is maximized. This approach is designed to sharpen the visibility of the $M_{\mathrm{LV}}$ trend, highlighting the impact of high redshift values. As illustrated in Fig. \ref{fig:mlv for six detectors}, the highest values of the Lorentz violation energy scale, $M_{\mathrm{LV}}$, are achieved for sources that are both nearest and of the lowest mass, as observed in the cases of aLIGO, CE, ET, and TianQin. Conversely, for Taiji and LISA, the trend deviates. Here, $M_{\rm LV}$ does not attain its maximum in regions characterized by lower total masses and smaller redshifts. This discrepancy arises because the innermost stable circular orbit frequency, $f_{\mathrm{ISCO}}$, in this domain, surpasses the upper-frequency limit of these detectors. This is indicated by Eq. (\ref{fisco}) and illustrated in Fig.~\ref{fig:fisco for Taiji and LISA}. Consequently, this suggests that Taiji and LISA might be less efficient in detecting the final inspiral phase in compact binary systems that have lower redshifts and smaller total masses.

The choice to initiate the redshift ($z$) analysis from 0.01 stems from the observation that the lowest redshift value recorded in the LVK event catalog is 0.01. Considering that the mass of a neutron star is typically around 1.5 $M_{\odot}$, we adopt this figure as the minimum mass threshold for our total mass range. We cap the redshift at $z=0.5$ and set the maximum total mass at 50 $M_{\odot}$, which corresponds to the median GW source mass reported in the LVK events. To streamline our analysis, we concentrate on binary systems comprising either two black holes (BBH) or two neutron stars (BNS), assuming equal mass for both components.

\section{CONCLUSION}\label{sec:5}
\renewcommand{\theequation}{5.\arabic{equation}} \setcounter{equation}{0}

With the advent of future detectors, GWs are poised to play a pivotal role in testing gravity in the strong field regime. Both ground-based and space-based detectors are designed to capture GWs across different frequency bands, spanning from $10^{-4}$ to $10^{4}$ Hz.  In this study, we delve into the investigation of the Lorentz-violating damping effect, which influences the propagation of GWs. We aim to evaluate the capability of both ground-based (aLIGO, CE, and ET) and space-based (Taiji, TianQin, and LISA) detectors in constraining this effect. We begin by formulating the modified equations of motion for the two polarizations of GWs. We then proceed to derive the altered GW waveform in the Fourier domain, incorporating the Lorentz-violating damping effect. Utilizing the FIM, we set out to quantify the constraints on the energy scale $M_{\mathrm{LV}}$, showcasing its projected sensitivity for each detector. For the FIM analysis, we establish detection thresholds for ground-based detectors and space-based detectors at $\rho > 8$ and $\rho > 15$, respectively. Additionally, we conduct a joint analysis of $M_{\mathrm{LV}}$ using simulated GW events to further our understanding of the constraints achievable with future GW observations.

For ground-based detectors, the tightest constraint from a single event is set by CE, with $M_{\rm LV} > 3.02 \times 10^{-21}$ GeV. When conducting a joint analysis of 100 GW events, this constraint improves to $M_{\rm LV} > 8.84 \times 10^{-21}$ GeV. Regarding space-based detectors, the results are in line with our expectations. For a single event, the constraints on $M_{\rm LV}$ from these detectors are approximately $\gtrsim 2 \times 10^{-24}\; {\rm GeV}$, which is roughly three orders of magnitude less stringent than those obtained from ground-based detectors. Upon performing a joint analysis of 100 simulated GW events, the constraints from the space-based detectors converge to approximately $\gtrsim 5 \times 10^{-24}\; {\rm GeV}$.

In our conservative estimation of event numbers, we note an improvement of more than twofold over the constraints obtained from individual events. We posit that, with the ongoing accumulation of observational data, even more stringent constraints on this effect will be achievable. Our analysis indicates that targeting the high-frequency band offers a more efficacious approach for constraining the Lorentz-violating damping effect. This suggests that ground-based detectors are more adept at imposing rigorous constraints on the effect in comparison to space-based detectors. Within this group of detectors, CE distinguishes itself by offering the most stringent lower bound on $M_{\mathrm{LV}}$.

Our analysis of the data distribution reveals that an effective strategy for aLIGO, CE, ET, and TianQin to constrain the Lorentz-violating damping effect involves concentrating on compact binary systems characterized by both a small total mass and low redshift. Conversely, for Taiji and LISA, targeting GW sources that have a small total mass but are situated at higher redshifts proves to be more appropriate. This strategic focus is informed by the differential sensitivity of these detectors to the frequency and amplitude of GW signals, which in turn affects their capability to place constraints on the Lorentz-violating effect.

\section*{ACKNOWLEDGMENTS}

We thank Hong-Chao Zhang, Chao Zhang, and Peng-Ju Wu for useful discussions. Tao Zhu and Bo-Yang Zhang are supported by the National Key Research and Development Program of China under Grant No.2020YFC2201503, the National Natural Science Foundation of China under Grants No.12275238 and No. 11675143, the Zhejiang Provincial Natural Science Foundation of China under Grants No.LR21A050001 and No. LY20A050002, and the Fundamental Research Funds for the Provincial Universities of Zhejiang in China under Grant No. RF-A2019015. Jing-Fei Zhang and Xin Zhang are supported by the National Natural Science Foundation of China (Grants Nos. 11975072, 11875102, and 11835009), the National SKA Program of China (Grants Nos. 2022SKA0110200 and 2022SKA0110203), and the National 111 Project (Grant No. B16009).

%\appendix

\section*{APPENDIX A: PARTIAL DERIVATIVES OF THE WAVEFORM OF BINARY INSPIRAL}
\renewcommand{\theequation}{A.\arabic{equation}} \setcounter{equation}{0}

In this appendix, we present the partial derivatives of the GW waveform parameters $ \{\ln \mathcal{A}, \ln \mathcal{M}, \ln \mathcal{\eta}, \phi_{\mathrm{c}}, t_{\mathrm{c}}, C_{\nu}\}$ as follows,
\begin{widetext}
\begin{eqnarray}
\frac{\partial \tilde{h}(f)}{\partial\ln{\cal{A}}}&=&\tilde{h}(f),\\
\frac{\partial \tilde{h}(f)}{\partial\ln \cal{M}}&=&\frac{5}{6} \tilde{h}(f),\\
\frac{\partial \tilde{h}(f)}{\partial\ln {\eta}}&=& \frac{1}{2} + i \eta \Bigg \{ -\frac{3}{128} \eta^{-2} u^{-5/3} C_{\Psi} + \frac{3}{128} \eta^{-1} u^{-5/3} \Bigg[ \frac{55}{9} u^{2/3} + \left(\frac{27145}{504} + \frac{3085}{36} \eta\right) u^{4/3} -  \frac{65\pi}{9} \left(1+\rm \ln{u}\right) u^{5/3}\nonumber\\
    &&+ \left(-\frac{15737765635}{3048192} + \frac{2255}{12} \pi^{2} + \frac{76055}{864} \eta - \frac{127825}{432} \eta^{2}\right) u ^{2} + \pi \left(\frac{378515}{1512} - \frac{74045}{378} \eta\right) u^{7/3} \Bigg]\Bigg \} , \label{A3}\\
    \frac{\partial \tilde{h}(f)}{\partial {\phi_{\rm c}}}&=&-i \tilde{h}(f),\\
    \frac{\partial \tilde{h}(f)}{\partial{t_{\rm c}}}&=&(2 \pi i f) \tilde{h}(f),\\
    \frac{\partial \tilde{h}(f)}{\partial{C_{\nu}}}&=&\Big [ \frac{1}{2} (2 \pi f)^{2} ((1+z)^{2} -1) \Big] \tilde{h}(f),
\end{eqnarray}
where $C_\Psi$ in Eq.~(\ref{A3}) is given by
\begin{eqnarray}
    C_{\Psi}&=&\left\{1+\left(\frac{3715}{756}+\frac{55}{9} \eta\right) u^{2 / 3}-16 \pi u\right.\nonumber\\
    && +\left(\frac{15293365}{508032}+\frac{27145}{504} \eta+\frac{3085}{72} \eta^2\right) u^{4 / 3}+\pi\left(\frac{38645}{756}-\frac{65}{9} \eta\right)(1+\ln u) u^{5 / 3}\nonumber \\
    && +\left[\frac{11583231236531}{4694215680}-\frac{640}{3} \pi^2-\frac{6848}{21} \gamma_{\mathrm{E}}-\frac{6848}{63} \ln (64 u)+\left(-\frac{15737765635}{3048192}+\frac{2255}{12} \pi^2\right) \eta\right.\nonumber \\
    && \left.\left.+\frac{76055}{1728} \eta^2-\frac{127825}{1296} \eta^3\right] u^2+\pi\left(\frac{77096675}{254016}+\frac{378515}{1512} \eta-\frac{74045}{756} \eta^2\right) u^{7 / 3}\right\}.
\end{eqnarray}
\end{widetext}


\begin{thebibliography}{399}

%\cite{LIGOScientific:2016aoc}
\bibitem{LIGOScientific:2016aoc}
B.~P.~Abbott \textit{et al.} [LIGO Scientific and Virgo],
Observation of Gravitational Waves from a Binary Black Hole Merger,
Phys. Rev. Lett. \textbf{116}, 061102 (2016)
%doi:10.1103/PhysRevLett.116.061102
[arXiv:1602.03837 [gr-qc]].
%10109 citations counted in INSPIRE as of 19 Oct 2023

%\cite{LIGOScientific:2018mvr}
\bibitem{LIGOScientific:2018mvr}
B.~P.~Abbott \textit{et al.} [LIGO Scientific and Virgo],
GWTC-1: A Gravitational-Wave Transient Catalog of Compact Binary Mergers Observed by LIGO and Virgo during the First and Second Observing Runs,
Phys. Rev. X \textbf{9}, no.3, 031040 (2019)
%doi:10.1103/PhysRevX.9.031040
[arXiv:1811.12907 [astro-ph.HE]].
%3118 citations counted in INSPIRE as of 19 Oct 2023

%\cite{LIGOScientific:2019lzm}
\bibitem{LIGOScientific:2019lzm}
R.~Abbott \textit{et al.} [LIGO Scientific and Virgo],
Open data from the first and second observing runs of Advanced LIGO and Advanced Virgo,
SoftwareX \textbf{13}, 100658 (2021)
%doi:10.1016/j.softx.2021.100658
[arXiv:1912.11716 [gr-qc]].
%399 citations counted in INSPIRE as of 19 Oct 2023

%\cite{LIGOScientific:2020aai}
\bibitem{LIGOScientific:2020aai}
B.~P.~Abbott \textit{et al.} [LIGO Scientific and Virgo],
GW190425: Observation of a Compact Binary Coalescence with Total Mass $\sim 3.4 M_{\odot}$,
Astrophys. J. Lett. \textbf{892}, no.1, L3 (2020)
%doi:10.3847/2041-8213/ab75f5
[arXiv:2001.01761 [astro-ph.HE]].
%1184 citations counted in INSPIRE as of 19 Oct 2023

%\cite{LIGOScientific:2021djp}
\bibitem{LIGOScientific:2021djp}
R.~Abbott \textit{et al.} [LIGO Scientific, VIRGO and KAGRA],
GWTC-3: Compact Binary Coalescences Observed by LIGO and Virgo During the Second Part of the Third Observing Run,
[arXiv:2111.03606 [gr-qc]].
%1285 citations counted in INSPIRE as of 19 Oct 2023

%\cite{Branchesi:2023mws}
\bibitem{Branchesi:2023mws}
M.~Branchesi, M.~Maggiore, D.~Alonso, C.~Badger, B.~Banerjee, F.~Beirnaert, E.~Belgacem, S.~Bhagwat, G.~Boileau and S.~Borhanian, \textit{et al.}
Science with the Einstein Telescope: a comparison of different designs,
JCAP \textbf{07}, 068 (2023)
%doi:10.1088/1475-7516/2023/07/068
[arXiv:2303.15923 [gr-qc]].
%45 citations counted in INSPIRE as of 21 Oct 2023

%\cite{Evans:2021gyd}
\bibitem{Evans:2021gyd}
M.~Evans, R.~X.~Adhikari, C.~Afle, S.~W.~Ballmer, S.~Biscoveanu, S.~Borhanian, D.~A.~Brown, Y.~Chen, R.~Eisenstein and A.~Gruson, \textit{et al.}
A Horizon Study for Cosmic Explorer: Science, Observatories, and Community,
[arXiv:2109.09882 [astro-ph.IM]].
%240 citations counted in INSPIRE as of 21 Oct 2023

%\cite{Ruan:2018tsw}
\bibitem{Ruan:2018tsw}
W.~H.~Ruan, Z.~K.~Guo, R.~G.~Cai and Y.~Z.~Zhang,
Taiji program: Gravitational-wave sources,
Int. J. Mod. Phys. A \textbf{35}, no.17, 2050075 (2020)
%doi:10.1142/S0217751X2050075X
[arXiv:1807.09495 [gr-qc]].
%390 citations counted in INSPIRE as of 06 Feb 2024

%\cite{Wu:2018clg}
\bibitem{Wu:2018clg}
Y.~L.~Wu,
Hyperunified field theory and Taiji program in space for GWD,
Int. J. Mod. Phys. A \textbf{33}, no.31, 1844014 (2018)
%doi:10.1142/S0217751X18440141
[arXiv:1805.10119 [physics.gen-ph]].
%12 citations counted in INSPIRE as of 06 Feb 2024

%\cite{Hu:2017mde}
\bibitem{Hu:2017mde}
W.~R.~Hu and Y.~L.~Wu,
The Taiji Program in Space for gravitational wave physics and the nature of gravity,
Natl. Sci. Rev. \textbf{4}, no.5, 685-686 (2017)
%doi:10.1093/nsr/nwx116
%411 citations counted in INSPIRE as of 06 Feb 2024

%\cite{Liu:2020eko}
\bibitem{Liu:2020eko}
S.~Liu, Y.~M.~Hu, J.~d.~Zhang and J.~Mei,
Science with the TianQin observatory: Preliminary results on stellar-mass binary black holes,
Phys. Rev. D \textbf{101}, no.10, 103027 (2020)
%doi:10.1103/PhysRevD.101.103027
[arXiv:2004.14242 [astro-ph.HE]].
%60 citations counted in INSPIRE as of 06 Feb 2024

%\cite{Wang:2019ryf}
\bibitem{Wang:2019ryf}
H.~T.~Wang, Z.~Jiang, A.~Sesana, E.~Barausse, S.~J.~Huang, Y.~F.~Wang, W.~F.~Feng, Y.~Wang, Y.~M.~Hu and J.~Mei, \textit{et al.}
Science with the TianQin observatory: Preliminary results on massive black hole binaries,
Phys. Rev. D \textbf{100}, no.4, 043003 (2019)
%doi:10.1103/PhysRevD.100.043003
[arXiv:1902.04423 [astro-ph.HE]].
%78 citations counted in INSPIRE as of 06 Feb 2024

%\cite{TianQin:2015yph}
\bibitem{TianQin:2015yph}
J.~Luo \textit{et al.} [TianQin],
TianQin: a space-borne gravitational wave detector,
Class. Quant. Grav. \textbf{33}, no.3, 035010 (2016)
%doi:10.1088/0264-9381/33/3/035010
[arXiv:1512.02076 [astro-ph.IM]].
%996 citations counted in INSPIRE as of 06 Feb 2024

%\cite{Luo:2020bls}
\bibitem{Luo:2020bls}
J.~Luo, Y.~Z.~Bai, L.~Cai, B.~Cao, W.~M.~Chen, Y.~Chen, D.~C.~Cheng, Y.~W.~Ding, H.~Z.~Duan and X.~Gou, \textit{et al.}
The first round result from the TianQin-1 satellite,
Class. Quant. Grav. \textbf{37}, no.18, 185013 (2020)
%doi:10.1088/1361-6382/aba66a
[arXiv:2008.09534 [physics.ins-det]].
%38 citations counted in INSPIRE as of 06 Feb 2024

%\cite{Milyukov:2020kyg}
\bibitem{Milyukov:2020kyg}
V.~K.~Milyukov,
TianQin Space-Based Gravitational Wave Detector: Key Technologies and Current State of Implementation,
Astron. Rep. \textbf{64}, no.12, 1067-1077 (2020)
%doi:10.1134/S1063772920120070
%9 citations counted in INSPIRE as of 06 Feb 2024

%\cite{Robson:2018ifk}
\bibitem{Robson:2018ifk}
T.~Robson, N.~J.~Cornish and C.~Liu,
The construction and use of LISA sensitivity curves,
Class. Quant. Grav. \textbf{36}, no.10, 105011 (2019)
%doi:10.1088/1361-6382/ab1101
[arXiv:1803.01944 [astro-ph.HE]].
%524 citations counted in INSPIRE as of 06 Feb 2024

%\cite{LISACosmologyWorkingGroup:2022jok}
\bibitem{LISACosmologyWorkingGroup:2022jok}
P.~Auclair \textit{et al.} [LISA Cosmology Working Group],
Cosmology with the Laser Interferometer Space Antenna,
Living Rev. Rel. \textbf{26}, no.1, 5 (2023)
%doi:10.1007/s41114-023-00045-2
[arXiv:2204.05434 [astro-ph.CO]].
%216 citations counted in INSPIRE as of 06 Feb 2024


%%%%%%
%\cite{LIGOScientific:2016lio}
\bibitem{LIGOScientific:2016lio}
B.~P.~Abbott \textit{et al.} [LIGO Scientific and Virgo],
Tests of general relativity with GW150914,
Phys. Rev. Lett. \textbf{116}, no.22, 221101 (2016)
[erratum: Phys. Rev. Lett. \textbf{121}, no.12, 129902 (2018)]
%doi:10.1103/PhysRevLett.116.221101
[arXiv:1602.03841 [gr-qc]].
%1595 citations counted in INSPIRE as of 19 Oct 2023

%\cite{LIGOScientific:2018dkp}
\bibitem{LIGOScientific:2018dkp}
B.~P.~Abbott \textit{et al.} [LIGO Scientific and Virgo],
Tests of General Relativity with GW170817,
Phys. Rev. Lett. \textbf{123}, no.1, 011102 (2019)
%doi:10.1103/PhysRevLett.123.011102
[arXiv:1811.00364 [gr-qc]].
%520 citations counted in INSPIRE as of 19 Oct 2023

%\cite{LIGOScientific:2017zic}
\bibitem{LIGOScientific:2017zic}
B.~P.~Abbott \textit{et al.} [LIGO Scientific, Virgo, Fermi-GBM and INTEGRAL],
Gravitational Waves and Gamma-rays from a Binary Neutron Star Merger: GW170817 and GRB 170817A,
Astrophys. J. Lett. \textbf{848}, no.2, L13 (2017)
%doi:10.3847/2041-8213/aa920c
[arXiv:1710.05834 [astro-ph.HE]].
%2686 citations counted in INSPIRE as of 19 Oct 2023

%\cite{LIGOScientific:2019fpa}
\bibitem{LIGOScientific:2019fpa}
B.~P.~Abbott \textit{et al.} [LIGO Scientific and Virgo],
Tests of General Relativity with the Binary Black Hole Signals from the LIGO-Virgo Catalog GWTC-1,
Phys. Rev. D \textbf{100}, no.10, 104036 (2019)
%doi:10.1103/PhysRevD.100.104036
[arXiv:1903.04467 [gr-qc]].
%657 citations counted in INSPIRE as of 19 Oct 2023

%\cite{LIGOScientific:2020tif}
\bibitem{LIGOScientific:2020tif}
R.~Abbott \textit{et al.} [LIGO Scientific and Virgo],
Tests of general relativity with binary black holes from the second LIGO-Virgo gravitational-wave transient catalog,
Phys. Rev. D \textbf{103}, no.12, 122002 (2021)
%doi:10.1103/PhysRevD.103.122002
[arXiv:2010.14529 [gr-qc]].
%510 citations counted in INSPIRE as of 19 Oct 2023

%\cite{Hoyle:2000cv}
\bibitem{Hoyle:2000cv}
C.~D.~Hoyle, U.~Schmidt, B.~R.~Heckel, E.~G.~Adelberger, J.~H.~Gundlach, D.~J.~Kapner and H.~E.~Swanson,
Submillimeter tests of the gravitational inverse square law: a search for 'large' extra dimensions,
Phys. Rev. Lett. \textbf{86}, 1418-1421 (2001)
%doi:10.1103/PhysRevLett.86.1418
[arXiv:hep-ph/0011014 [hep-ph]].
%542 citations counted in INSPIRE as of 18 Oct 2023

%\cite{Everitt:2011hp}
\bibitem{Everitt:2011hp}
C.~W.~F.~Everitt, D.~B.~DeBra, B.~W.~Parkinson, J.~P.~Turneaure, J.~W.~Conklin, M.~I.~Heifetz, G.~M.~Keiser, A.~S.~Silbergleit, T.~Holmes and J.~Kolodziejczak, \textit{et al.}
Gravity Probe B: Final Results of a Space Experiment to Test General Relativity,
Phys. Rev. Lett. \textbf{106}, 221101 (2011)
%doi:10.1103/PhysRevLett.106.221101
[arXiv:1105.3456 [gr-qc]].
%555 citations counted in INSPIRE as of 18 Oct 2023

%\cite{Sabulsky:2018jma}
\bibitem{Sabulsky:2018jma}
D.~O.~Sabulsky, I.~Dutta, E.~A.~Hinds, B.~Elder, C.~Burrage and E.~J.~Copeland,
Experiment to detect dark energy forces using atom interferometry,
Phys. Rev. Lett. \textbf{123}, no.6, 061102 (2019)
%doi:10.1103/PhysRevLett.123.061102
[arXiv:1812.08244 [physics.atom-ph]].
%85 citations counted in INSPIRE as of 18 Oct 2023

%\cite{LIGOScientific:2020iuh}
\bibitem{LIGOScientific:2020iuh}
R.~Abbott \textit{et al.} [LIGO Scientific and Virgo],
GW190521: A Binary Black Hole Merger with a Total Mass of $150  M_{\odot}$,
Phys. Rev. Lett. \textbf{125}, no.10, 101102 (2020)
%doi:10.1103/PhysRevLett.125.101102
[arXiv:2009.01075 [gr-qc]].
%931 citations counted in INSPIRE as of 19 Oct 2023

%\cite{LIGOScientific:2017vwq}
\bibitem{LIGOScientific:2017vwq}
B.~P.~Abbott \textit{et al.} [LIGO Scientific and Virgo],
GW170817: Observation of Gravitational Waves from a Binary Neutron Star Inspiral,
Phys. Rev. Lett. \textbf{119}, no.16, 161101 (2017)
%doi:10.1103/PhysRevLett.119.161101
[arXiv:1710.05832 [gr-qc]].
%7330 citations counted in INSPIRE as of 19 Oct 2023

%\cite{LIGOScientific:2020ibl}
\bibitem{LIGOScientific:2020ibl}
R.~Abbott \textit{et al.} [LIGO Scientific and Virgo],
GWTC-2: Compact Binary Coalescences Observed by LIGO and Virgo During the First Half of the Third Observing Run,
Phys. Rev. X \textbf{11}, 021053 (2021)
%doi:10.1103/PhysRevX.11.021053
[arXiv:2010.14527 [gr-qc]].
%1722 citations counted in INSPIRE as of 19 Oct 2023

%\cite{Cognola:2006eg}
\bibitem{Cognola:2006eg}
G.~Cognola, E.~Elizalde, S.~Nojiri, S.~D.~Odintsov and S.~Zerbini,
Dark energy in modified Gauss-Bonnet gravity: Late-time acceleration and the hierarchy problem,
Phys. Rev. D \textbf{73}, 084007 (2006)
%doi:10.1103/PhysRevD.73.084007
[arXiv:hep-th/0601008 [hep-th]].
%696 citations counted in INSPIRE as of 19 Oct 2023

%\cite{Copeland:2006wr}
\bibitem{Copeland:2006wr}
E.~J.~Copeland, M.~Sami and S.~Tsujikawa,
Dynamics of dark energy,
Int. J. Mod. Phys. D \textbf{15}, 1753-1936 (2006)
%doi:10.1142/S021827180600942X
[arXiv:hep-th/0603057 [hep-th]].
%5330 citations counted in INSPIRE as of 19 Oct 2023

%\cite{Frieman:2008sn}
\bibitem{Frieman:2008sn}
J.~Frieman, M.~Turner and D.~Huterer,
Dark Energy and the Accelerating Universe,
Ann. Rev. Astron. Astrophys. \textbf{46}, 385-432 (2008)
%doi:10.1146/annurev.astro.46.060407.145243
[arXiv:0803.0982 [astro-ph]].
%1418 citations counted in INSPIRE as of 19 Oct 2023

%\cite{Jackiw:2003pm}
\bibitem{Jackiw:2003pm}
R.~Jackiw and S.~Y.~Pi,
Chern-Simons modification of general relativity,
Phys. Rev. D \textbf{68}, 104012 (2003)
%doi:10.1103/PhysRevD.68.104012
[arXiv:gr-qc/0308071 [gr-qc]].
%653 citations counted in INSPIRE as of 12 Mar 2024

%\cite{Yunes:2010yf}
\bibitem{Yunes:2010yf}
N.~Yunes, R.~O'Shaughnessy, B.~J.~Owen and S.~Alexander,
Testing gravitational parity violation with coincident gravitational waves and short gamma-ray bursts,
Phys. Rev. D \textbf{82}, 064017 (2010)
%doi:10.1103/PhysRevD.82.064017
[arXiv:1005.3310 [gr-qc]].
%96 citations counted in INSPIRE as of 12 Mar 2024

%\cite{Yagi:2012vf}
\bibitem{Yagi:2012vf}
K.~Yagi, N.~Yunes and T.~Tanaka,
Gravitational Waves from Quasi-Circular Black Hole Binaries in Dynamical Chern-Simons Gravity,
Phys. Rev. Lett. \textbf{109}, 251105 (2012)
[erratum: Phys. Rev. Lett. \textbf{116}, no.16, 169902 (2016); erratum: Phys. Rev. Lett. \textbf{124}, no.2, 029901 (2020)]
%doi:10.1103/PhysRevLett.116.169902
[arXiv:1208.5102 [gr-qc]].
%121 citations counted in INSPIRE as of 12 Mar 2024

%\cite{Wang:2017brl}
\bibitem{Wang:2017brl}
A.~Wang,
Ho\v{r}ava gravity at a Lifshitz point: A progress report,
Int. J. Mod. Phys. D \textbf{26}, no.07, 1730014 (2017)
%doi:10.1142/S0218271817300142
[arXiv:1701.06087 [gr-qc]].
%169 citations counted in INSPIRE as of 12 Mar 2024

%\cite{Sefiedgar:2010we}
\bibitem{Sefiedgar:2010we}
A.~S.~Sefiedgar, K.~Nozari and H.~R.~Sepangi,
Modified dispersion relations in extra dimensions,
Phys. Lett. B \textbf{696}, 119-123 (2011)
%doi:10.1016/j.physletb.2010.11.067
[arXiv:1012.1406 [gr-qc]].
%59 citations counted in INSPIRE as of 12 Mar 2024

%\cite{Crisostomi:2017ugk}
\bibitem{Crisostomi:2017ugk}
M.~Crisostomi, K.~Noui, C.~Charmousis and D.~Langlois,
Beyond Lovelock gravity: Higher derivative metric theories,
Phys. Rev. D \textbf{97}, no.4, 044034 (2018)
%doi:10.1103/PhysRevD.97.044034
[arXiv:1710.04531 [hep-th]].
%103 citations counted in INSPIRE as of 12 Mar 2024

%\cite{Gao:2014soa}
\bibitem{Gao:2014soa}
X.~Gao,
Unifying framework for scalar-tensor theories of gravity,
Phys. Rev. D \textbf{90}, 081501 (2014)
%doi:10.1103/PhysRevD.90.081501
[arXiv:1406.0822 [gr-qc]].
%171 citations counted in INSPIRE as of 12 Mar 2024

%\cite{Gao:2019lpz}
\bibitem{Gao:2019lpz}
X.~Gao, C.~Kang and Z.~B.~Yao,
Spatially Covariant Gravity: Perturbative Analysis and Field Transformations,
Phys. Rev. D \textbf{99}, no.10, 104015 (2019)
%doi:10.1103/PhysRevD.99.104015
[arXiv:1902.07702 [gr-qc]].
%22 citations counted in INSPIRE as of 12 Mar 2024

%\cite{Jacobson:2000xp}
\bibitem{Jacobson:2000xp}
T.~Jacobson and D.~Mattingly,
Gravity with a dynamical preferred frame,
Phys. Rev. D \textbf{64}, 024028 (2001)
%doi:10.1103/PhysRevD.64.024028
[arXiv:gr-qc/0007031 [gr-qc]].
%812 citations counted in INSPIRE as of 21 Oct 2023

%\cite{Eling:2004dk}
\bibitem{Eling:2004dk}
C.~Eling, T.~Jacobson and D.~Mattingly,
Einstein-Aether theory,
[arXiv:gr-qc/0410001 [gr-qc]].
%170 citations counted in INSPIRE as of 21 Oct 2023

%\cite{Jacobson:2007veq}
\bibitem{Jacobson:2007veq}
T.~Jacobson,
Einstein-aether gravity: A Status report,
PoS \textbf{QG-PH}, 020 (2007)
%doi:10.22323/1.043.0020
[arXiv:0801.1547 [gr-qc]].
%329 citations counted in INSPIRE as of 21 Oct 2023

%\cite{Li:2007vz}
\bibitem{Li:2007vz}
B.~Li, D.~Fonseca Mota and J.~D.~Barrow,
Detecting a Lorentz-Violating Field in Cosmology,
Phys. Rev. D \textbf{77}, 024032 (2008)
%doi:10.1103/PhysRevD.77.024032
[arXiv:0709.4581 [astro-ph]].
%87 citations counted in INSPIRE as of 21 Oct 2023

%\cite{Battye:2017zvv}
\bibitem{Battye:2017zvv}
R.~A.~Battye, F.~Pace and D.~Trinh,
Cosmological perturbation theory in Generalized Einstein-Aether models,
Phys. Rev. D \textbf{96}, no.6, 064041 (2017)
%doi:10.1103/PhysRevD.96.064041
[arXiv:1707.06508 [astro-ph.CO]].
%26 citations counted in INSPIRE as of 21 Oct 2023

%\cite{Zhang:2023kzs}
\bibitem{Zhang:2023kzs}
C.~Zhang, A.~Wang and T.~Zhu,
Odd-parity perturbations of the wormhole-like geometries and quasi-normal modes in Einstein-\AE{}ther theory,
JCAP \textbf{05}, 059 (2023)
%doi:10.1088/1475-7516/2023/05/059
[arXiv:2303.08399 [gr-qc]].
%0 citations counted in INSPIRE as of 07 Feb 2024

%\cite{Liu:2021yev}
\bibitem{Liu:2021yev}
C.~Liu, S.~Yang, Q.~Wu and T.~Zhu,
Thin accretion disk onto slowly rotating black holes in Einstein-\AE{}ther theory,
JCAP \textbf{02}, no.02, 034 (2022)
%doi:10.1088/1475-7516/2022/02/034
[arXiv:2107.04811 [gr-qc]].
%27 citations counted in INSPIRE as of 07 Feb 2024

%\cite{Zhu:2019ura}
\bibitem{Zhu:2019ura}
T.~Zhu, Q.~Wu, M.~Jamil and K.~Jusufi,
Shadows and deflection angle of charged and slowly rotating black holes in Einstein-\AE{}ther theory,
Phys. Rev. D \textbf{100}, no.4, 044055 (2019)
%doi:10.1103/PhysRevD.100.044055
[arXiv:1906.05673 [gr-qc]].
%122 citations counted in INSPIRE as of 07 Feb 2024


%\cite{Horava:2009uw}
\bibitem{Horava:2009uw}
P.~Horava,
Quantum Gravity at a Lifshitz Point,
Phys. Rev. D \textbf{79}, 084008 (2009)
%doi:10.1103/PhysRevD.79.084008
[arXiv:0901.3775 [hep-th]].
%2389 citations counted in INSPIRE as of 20 Oct 2023

%\cite{Takahashi:2009wc}
\bibitem{Takahashi:2009wc}
T.~Takahashi and J.~Soda,
Chiral Primordial Gravitational Waves from a Lifshitz Point,
Phys. Rev. Lett. \textbf{102}, 231301 (2009)
%doi:10.1103/PhysRevLett.102.231301
[arXiv:0904.0554 [hep-th]].
%300 citations counted in INSPIRE as of 20 Oct 2023

%\cite{Wang:2012fi}
\bibitem{Wang:2012fi}
A.~Wang, Q.~Wu, W.~Zhao and T.~Zhu,
Polarizing primordial gravitational waves by parity violation,
Phys. Rev. D \textbf{87}, no.10, 103512 (2013)
%doi:10.1103/PhysRevD.87.103512
[arXiv:1208.5490 [astro-ph.CO]].
%69 citations counted in INSPIRE as of 20 Oct 2023

%\cite{Zhu:2013fja}
\bibitem{Zhu:2013fja}
T.~Zhu, W.~Zhao, Y.~Huang, A.~Wang and Q.~Wu,
Effects of parity violation on non-gaussianity of primordial gravitational waves in Ho\v{r}ava-Lifshitz gravity,
Phys. Rev. D \textbf{88}, 063508 (2013)
%doi:10.1103/PhysRevD.88.063508
[arXiv:1305.0600 [hep-th]].
%67 citations counted in INSPIRE as of 20 Oct 2023

%\cite{Gao:2020qxy}
\bibitem{Gao:2020qxy}
X.~Gao,
Higher derivative scalar-tensor theory from the spatially covariant gravity: a linear algebraic analysis,
JCAP \textbf{11}, 004 (2020)
%doi:10.1088/1475-7516/2020/11/004
[arXiv:2006.15633 [gr-qc]].
%8 citations counted in INSPIRE as of 21 Oct 2023

%\cite{Gao:2020yzr}
\bibitem{Gao:2020yzr}
X.~Gao and Y.~M.~Hu,
Higher derivative scalar-tensor theory and spatially covariant gravity: the correspondence,
Phys. Rev. D \textbf{102}, 084006 (2020)
%doi:10.1103/PhysRevD.102.084006
[arXiv:2004.07752 [gr-qc]].
%16 citations counted in INSPIRE as of 21 Oct 2023

%\cite{Gao:2019twq}
\bibitem{Gao:2019twq}
X.~Gao and Z.~B.~Yao, 
Spatially covariant gravity theories with two tensorial degrees of freedom: the formalism,
Phys. Rev. D \textbf{101}, 064018 (2020)
%doi:10.1103/PhysRevD.101.064018
[arXiv:1910.13995 [gr-qc]].
%29 citations counted in INSPIRE as of 21 Oct 2023

%42
%\cite{Joshi:2021azw}
\bibitem{Joshi:2021azw}
P.~Joshi and S.~Panda,
Higher derivative scalar tensor theory in unitary gauge,
JCAP \textbf{03}, 022 (2022)
%doi:10.1088/1475-7516/2022/03/022
[arXiv:2111.11791 [hep-th]].
%6 citations counted in INSPIRE as of 21 Oct 2023

\bibitem{Gao:2019liu}
X.~Gao and X.~Y.~Hong,
Propagation of gravitational waves in a cosmological background,
Phys. Rev. D \textbf{101}, no.6, 064057 (2020)
%doi:10.1103/PhysRevD.101.064057
[arXiv:1906.07131 [gr-qc]].

%add
%\cite{Kostelecky:2016kfm}
\bibitem{Kostelecky:2016kfm}
V.~A.~Kosteleck\'y and M.~Mewes,
Testing local Lorentz invariance with gravitational waves,
Phys. Lett. B \textbf{757}, 510-514 (2016)
%doi:10.1016/j.physletb.2016.04.040
[arXiv:1602.04782 [gr-qc]].
%177 citations counted in INSPIRE as of 21 Oct 2023

%\cite{Bailey:2006fd}
\bibitem{Bailey:2006fd}
Q.~G.~Bailey and V.~A.~Kostelecky,
Signals for Lorentz violation in post-Newtonian gravity,
Phys. Rev. D \textbf{74}, 045001 (2006)
%doi:10.1103/PhysRevD.74.045001
[arXiv:gr-qc/0603030 [gr-qc]].
%419 citations counted in INSPIRE as of 21 Oct 2023

%\cite{Mewes:2019dhj}
\bibitem{Mewes:2019dhj}
M.~Mewes,
Signals for Lorentz violation in gravitational waves,
Phys. Rev. D \textbf{99}, 104062 (2019)
%doi:10.1103/PhysRevD.99.104062
[arXiv:1905.00409 [gr-qc]].
%72 citations counted in INSPIRE as of 21 Oct 2023

%\cite{Shao:2020shv}
\bibitem{Shao:2020shv}
L.~Shao,
Combined search for anisotropic birefringence in the gravitational-wave transient catalog GWTC-1,
Phys. Rev. D \textbf{101}, 104019 (2020)
%doi:10.1103/PhysRevD.101.104019
[arXiv:2002.01185 [hep-ph]].
%44 citations counted in INSPIRE as of 21 Oct 2023

%\cite{Wang:2021ctl}
\bibitem{Wang:2021ctl}
Z.~Wang, L.~Shao and C.~Liu,
New Limits on the Lorentz/CPT Symmetry Through 50 Gravitational-wave Events,
Astrophys. J. \textbf{921}, 158 (2021)
%doi:10.3847/1538-4357/ac223c
[arXiv:2108.02974 [gr-qc]].
%24 citations counted in INSPIRE as of 21 Oct 2023

%\cite{Kostelecky:2017zob}
\bibitem{Kostelecky:2017zob}
V.~A.~Kosteleck\'y and M.~Mewes,
Lorentz and Diffeomorphism Violations in Linearized Gravity,
Phys. Lett. B \textbf{779}, 136-142 (2018)
%doi:10.1016/j.physletb.2018.01.082
[arXiv:1712.10268 [gr-qc]].
%64 citations counted in INSPIRE as of 21 Oct 2023

%\cite{Niu:2022yhr}
\bibitem{Niu:2022yhr}
R.~Niu, T.~Zhu and W.~Zhao,
Testing Lorentz invariance of gravity in the Standard-Model Extension with GWTC-3,
JCAP \textbf{12}, 011 (2022)
%doi:10.1088/1475-7516/2022/12/011
[arXiv:2202.05092 [gr-qc]].
%22 citations counted in INSPIRE as of 07 Feb 2024

%44
\bibitem{Colombo:2014lta}
M.~Colombo, A.~E.~Gumrukcuoglu and T.~P.~Sotiriou,
Ho\v{r}ava gravity with mixed derivative terms,
Phys. Rev. D \textbf{91}, no.4, 044021 (2015)
%doi:10.1103/PhysRevD.91.044021
[arXiv:1410.6360 [hep-th]].

%45
%\cite{Zhu:2022uoq}
\bibitem{Zhu:2022uoq}
T.~Zhu, W.~Zhao and A.~Wang,
Gravitational wave constraints on spatial covariant gravities,
Phys. Rev. D \textbf{107}, no.4, 044051 (2023)
%doi:10.1103/PhysRevD.107.044051
[arXiv:2211.04711 [gr-qc]].
%9 citations counted in INSPIRE as of 28 Aug 2023

\bibitem{Ghosh:2023xes}
R.~Ghosh, S.~Nair, L.~Pathak, S.~Sarkar and A.~S.~Sengupta,
Does the speed of gravitational waves depend on the source velocity?,
Phys. Rev. D \textbf{108}, 124017 (2023)
%doi:10.1103/PhysRevD.108.124017
[arXiv:2304.14820 [gr-qc]].

%46
%\cite{Gong:2021jgg}
\bibitem{Gong:2021jgg}
C.~Gong, T.~Zhu, R.~Niu, Q.~Wu, J.~L.~Cui, X.~Zhang, W.~Zhao and A.~Wang,
Gravitational wave constraints on Lorentz and parity violations in gravity: High-order spatial derivative cases,
Phys. Rev. D \textbf{105}, 044034 (2022)
%doi:10.1103/PhysRevD.105.044034
[arXiv:2112.06446 [gr-qc]].
%26 citations counted in INSPIRE as of 23 Oct 2023

%47
%\cite{Mirshekari:2011yq}
\bibitem{Mirshekari:2011yq}
S.~Mirshekari, N.~Yunes and C.~M.~Will,
Constraining Generic Lorentz Violation and the Speed of the Graviton with Gravitational Waves,
Phys. Rev. D \textbf{85}, 024041 (2012)
%doi:10.1103/PhysRevD.85.024041
[arXiv:1110.2720 [gr-qc]].
%180 citations counted in INSPIRE as of 24 Oct 2023

%48
%\cite{Will:1997bb}
\bibitem{Will:1997bb}
C.~M.~Will,
Bounding the mass of the graviton using gravitational wave observations of inspiralling compact binaries,
Phys. Rev. D \textbf{57}, 2061-2068 (1998)
%doi:10.1103/PhysRevD.57.2061
[arXiv:gr-qc/9709011 [gr-qc]].
%354 citations counted in INSPIRE as of 24 Oct 2023

%\cite{Haegel:2022ymk}
\bibitem{Haegel:2022ymk}
L.~Haegel, K.~O'Neal-Ault, Q.~G.~Bailey, J.~D.~Tasson, M.~Bloom and L.~Shao,
Search for anisotropic, birefringent spacetime-symmetry breaking in gravitational wave propagation from GWTC-3,
Phys. Rev. D \textbf{107}, no.6, 6 (2023)
%doi:10.1103/PhysRevD.107.064031
[arXiv:2210.04481 [gr-qc]].
%20 citations counted in INSPIRE as of 13 Mar 2024

\bibitem{Zhu:2023wci}
T.~Zhu, W.~Zhao, J.~M.~Yan, C.~Gong and A.~Wang,
Tests of modified gravitational wave propagations with gravitational waves,
[arXiv:2304.09025 [gr-qc]].

%\cite{Yu:2023ico}
\bibitem{Yu:2023ico}
J.~Yu, Z.~Liu, X.~Yang, Y.~Wang, P.~Zhang, X.~Zhang and W.~Zhao,
Measuring the Hubble Constant of Binary Neutron Star and Neutron Star\textendash{}Black Hole Coalescences: Bright Sirens and Dark Sirens,
Astrophys. J. Suppl. \textbf{270}, no.2, 24 (2024)
%doi:10.3847/1538-4365/ad0ece
[arXiv:2311.11588 [astro-ph.HE]].
%1 citations counted in INSPIRE as of 05 Feb 2024

%\cite{Jin:2023tou}
\bibitem{Jin:2023tou}
S.~J.~Jin, R.~Q.~Zhu, J.~Y.~Song, T.~Han, J.~F.~Zhang and X.~Zhang,
Standard siren cosmology in the era of the 2.5-generation ground-based gravitational wave detectors: bright and dark sirens of LIGO Voyager and NEMO,
[arXiv:2309.11900 [astro-ph.CO]].
%4 citations counted in INSPIRE as of 05 Feb 2024

%\cite{Shao:2023agv}
\bibitem{Shao:2023agv}
Y.~Shao, Y.~Xu, Y.~Wang, W.~Yang, R.~Li, X.~Zhang and X.~Chen,
The 21-cm forest as a simultaneous probe of dark matter and cosmic heating history,
Nature Astron. \textbf{7}, no.9, 1116-1126 (2023)
%doi:10.1038/s41550-023-02024-7
[arXiv:2307.04130 [astro-ph.CO]].
%2 citations counted in INSPIRE as of 05 Feb 2024

%\cite{Jin:2023sfc}
\bibitem{Jin:2023sfc}
S.~J.~Jin, Y.~Z.~Zhang, J.~Y.~Song, J.~F.~Zhang and X.~Zhang,
Taiji-TianQin-LISA network: Precisely measuring the Hubble constant using both bright and dark sirens,
Sci. China Phys. Mech. Astron. \textbf{67}, no.2, 220412 (2024)
%doi:10.1007/s11433-023-2276-1
[arXiv:2305.19714 [astro-ph.CO]].
%9 citations counted in INSPIRE as of 05 Feb 2024

%\cite{Zhang:2023gaz}
\bibitem{Zhang:2023gaz}
M.~Zhang, Y.~Li, J.~F.~Zhang and X.~Zhang,
H\,i intensity mapping with MeerKAT: forecast for delay power spectrum measurement using interferometer mode,
Mon. Not. Roy. Astron. Soc. \textbf{524}, no.2, 2420-2430 (2023)
%doi:10.1093/mnras/stad2033
[arXiv:2301.04445 [astro-ph.CO]].
%4 citations counted in INSPIRE as of 05 Feb 2024

%\cite{Wu:2022jkf}
\bibitem{Wu:2022jkf}
P.~J.~Wu, Y.~Li, J.~F.~Zhang and X.~Zhang,
Prospects for measuring dark energy with 21 cm intensity mapping experiments: A joint survey strategy,
Sci. China Phys. Mech. Astron. \textbf{66}, no.7, 270413 (2023)
%doi:10.1007/s11433-022-2104-7
[arXiv:2212.07681 [astro-ph.CO]].
%11 citations counted in INSPIRE as of 05 Feb 2024

%\cite{Song:2022siz}
\bibitem{Song:2022siz}
J.~Y.~Song, L.~F.~Wang, Y.~Li, Z.~W.~Zhao, J.~F.~Zhang, W.~Zhao and X.~Zhang,
Synergy between CSST galaxy survey and gravitational-wave observation: Inferring the Hubble constant from dark standard sirens,
Sci. China Phys. Mech. Astron. \textbf{67}, no.3, 230411 (2024)
%doi:10.1007/s11433-023-2260-2
[arXiv:2212.00531 [astro-ph.CO]].
%13 citations counted in INSPIRE as of 05 Feb 2024

%\cite{Jin:2022qnj}
\bibitem{Jin:2022qnj}
S.~J.~Jin, T.~N.~Li, J.~F.~Zhang and X.~Zhang,
Prospects for measuring the Hubble constant and dark energy using gravitational-wave dark sirens with neutron star tidal deformation,
JCAP \textbf{08}, 070 (2023)
%doi:10.1088/1475-7516/2023/08/070
[arXiv:2202.11882 [gr-qc]].
%26 citations counted in INSPIRE as of 05 Feb 2024

%\cite{Wu:2022dgy}
\bibitem{Wu:2022dgy}
P.~J.~Wu, Y.~Shao, S.~J.~Jin and X.~Zhang,
A path to precision cosmology: synergy between four promising late-universe cosmological probes,
JCAP \textbf{06}, 052 (2023)
%doi:10.1088/1475-7516/2023/06/052
[arXiv:2202.09726 [astro-ph.CO]].
%22 citations counted in INSPIRE as of 05 Feb 2024

%\cite{Wang:2022oou}
\bibitem{Wang:2022oou}
L.~F.~Wang, Y.~Shao, J.~F.~Zhang and X.~Zhang,
Ultra-low-frequency gravitational waves from individual supermassive black hole binaries as standard sirens,
[arXiv:2201.00607 [astro-ph.CO]].
%16 citations counted in INSPIRE as of 05 Feb 2024

%\cite{Wu:2021vfz}
\bibitem{Wu:2021vfz}
P.~J.~Wu and X.~Zhang,
Prospects for measuring dark energy with 21 cm intensity mapping experiments,
JCAP \textbf{01}, no.01, 060 (2022)
%doi:10.1088/1475-7516/2022/01/060
[arXiv:2108.03552 [astro-ph.CO]].
%12 citations counted in INSPIRE as of 06 Feb 2024

%\cite{Jin:2021pcv}
\bibitem{Jin:2021pcv}
S.~J.~Jin, L.~F.~Wang, P.~J.~Wu, J.~F.~Zhang and X.~Zhang,
How can gravitational-wave standard sirens and 21-cm intensity mapping jointly provide a precise late-universe cosmological probe?,
Phys. Rev. D \textbf{104}, no.10, 103507 (2021)
%doi:10.1103/PhysRevD.104.103507
[arXiv:2106.01859 [astro-ph.CO]].
%31 citations counted in INSPIRE as of 06 Feb 2024

%\cite{Zhang:2021yof}
\bibitem{Zhang:2021yof}
M.~Zhang, B.~Wang, P.~J.~Wu, J.~Z.~Qi, Y.~Xu, J.~F.~Zhang and X.~Zhang,
Prospects for Constraining Interacting Dark Energy Models with 21 cm Intensity Mapping Experiments,
Astrophys. J. \textbf{918}, no.2, 56 (2021)
%doi:10.3847/1538-4357/ac0ef5
[arXiv:2102.03979 [astro-ph.CO]].
%26 citations counted in INSPIRE as of 06 Feb 2024

%\cite{Yu:2021nvx}
\bibitem{Yu:2021nvx}
J.~Yu, H.~Song, S.~Ai, H.~Gao, F.~Wang, Y.~Wang, Y.~Lu, W.~Fang and W.~Zhao,
Multimessenger Detection Rates and Distributions of Binary Neutron Star Mergers and Their Cosmological Implications,
Astrophys. J. \textbf{916}, no.1, 54 (2021)
%doi:10.3847/1538-4357/ac0628
[arXiv:2104.12374 [astro-ph.HE]].
%30 citations counted in INSPIRE as of 06 Feb 2024

%\cite{Zhao:2019gyk}
\bibitem{Zhao:2019gyk}
Z.~W.~Zhao, L.~F.~Wang, J.~F.~Zhang and X.~Zhang,
Prospects for improving cosmological parameter estimation with gravitational-wave standard sirens from Taiji,
Sci. Bull. \textbf{65}, no.16, 1340-1348 (2020)
%doi:10.1016/j.scib.2020.04.032
[arXiv:1912.11629 [astro-ph.CO]].
%36 citations counted in INSPIRE as of 06 Feb 2024

%\cite{Zhang:2019loq}
\bibitem{Zhang:2019loq}
J.~F.~Zhang, M.~Zhang, S.~J.~Jin, J.~Z.~Qi and X.~Zhang,
Cosmological parameter estimation with future gravitational wave standard siren observation from the Einstein Telescope,
JCAP \textbf{09}, 068 (2019)
%doi:10.1088/1475-7516/2019/09/068
[arXiv:1907.03238 [astro-ph.CO]].
%43 citations counted in INSPIRE as of 06 Feb 2024

%\cite{Wang:2019tto}
\bibitem{Wang:2019tto}
L.~F.~Wang, Z.~W.~Zhao, J.~F.~Zhang and X.~Zhang,
A preliminary forecast for cosmological parameter estimation with gravitational-wave standard sirens from TianQin,
JCAP \textbf{11}, 012 (2020)
%doi:10.1088/1475-7516/2020/11/012
[arXiv:1907.01838 [astro-ph.CO]].
%36 citations counted in INSPIRE as of 06 Feb 2024

%\cite{Zhang:2019ple}
\bibitem{Zhang:2019ple}
J.~F.~Zhang, H.~Y.~Dong, J.~Z.~Qi and X.~Zhang,
Prospect for constraining holographic dark energy with gravitational wave standard sirens from the Einstein Telescope,
Eur. Phys. J. C \textbf{80}, no.3, 217 (2020)
%doi:10.1140/epjc/s10052-020-7767-3
[arXiv:1906.07504 [astro-ph.CO]].
%36 citations counted in INSPIRE as of 06 Feb 2024

%\cite{Zhao:2019xmm}
\bibitem{Zhao:2019xmm}
W.~Zhao, T.~Zhu, J.~Qiao and A.~Wang,
Waveform of gravitational waves in the general parity-violating gravities,
Phys. Rev. D \textbf{101}, no.2, 024002 (2020)
%doi:10.1103/PhysRevD.101.024002
[arXiv:1909.10887 [gr-qc]].
%73 citations counted in INSPIRE as of 05 Feb 2024

%91
\bibitem{Lagos:2019kds}
M.~Lagos, M.~Fishbach, P.~Landry and D.~E.~Holz,
Standard sirens with a running Planck mass,
Phys. Rev. D \textbf{99}, no.8, 083504 (2019)
%doi:10.1103/PhysRevD.99.083504
[arXiv:1901.03321 [astro-ph.CO]].


\bibitem{ADM}
 C.W. Misner, K.S. Thorne, and J.A. Wheeler, Gravi- tation (W.H. Freeman and Company, San Francisco, 1973), pp.484-528.

\bibitem{Qiao:2019wsh}
J.~Qiao, T.~Zhu, W.~Zhao and A.~Wang,
Waveform of gravitational waves in the ghost-free parity-violating gravities,
Phys. Rev. D \textbf{100}, 124058 (2019)
%doi:10.1103/PhysRevD.100.124058
[arXiv:1909.03815 [gr-qc]].

\bibitem{Jaranowski:1998qm}
P.~Jaranowski, A.~Krolak and B.~F.~Schutz,
Data analysis of gravitational-wave signals from spinning neutron stars. 1. The Signal and its detection,
Phys. Rev. D \textbf{58}, 063001 (1998)
%doi:10.1103/PhysRevD.58.063001
[arXiv:gr-qc/9804014 [gr-qc]].

%53
\bibitem{Zhao:2017cbb}
W.~Zhao and L.~Wen,
Localization accuracy of compact binary coalescences detected by the third-generation gravitational-wave detectors and implication for cosmology,
Phys. Rev. D \textbf{97}, no.6, 064031 (2018)
%doi:10.1103/PhysRevD.97.064031
[arXiv:1710.05325 [astro-ph.CO]].

%54
%\cite{Wang:2022yxb}
\bibitem{Wang:2022yxb}
Z.~Wang, J.~Zhao, Z.~An, L.~Shao and Z.~Cao,
Simultaneous bounds on the gravitational dipole radiation and varying gravitational constant from compact binary inspirals,
Phys. Lett. B \textbf{834}, 137416 (2022)
%doi:10.1016/j.physletb.2022.137416
[arXiv:2208.11913 [gr-qc]].
%5 citations counted in INSPIRE as of 24 Oct 2023

%55
%\cite{KAGRA:2023pio}
\bibitem{KAGRA:2023pio}
R.~Abbott \textit{et al.} [KAGRA, VIRGO and LIGO Scientific],
Open Data from the Third Observing Run of LIGO, Virgo, KAGRA, and GEO,
Astrophys. J. Suppl. \textbf{267}, 29 (2023)
%doi:10.3847/1538-4365/acdc9f
[arXiv:2302.03676 [gr-qc]].
%71 citations counted in INSPIRE as of 24 Nov 2023

%56
%\cite{Hild:2010id}
\bibitem{Hild:2010id}
S.~Hild, M.~Abernathy, F.~Acernese, P.~Amaro-Seoane, N.~Andersson, K.~Arun, F.~Barone, B.~Barr, M.~Barsuglia and M.~Beker, \textit{et al.}
Sensitivity Studies for Third-Generation Gravitational Wave Observatories,
Class. Quant. Grav. \textbf{28}, 094013 (2011)
%doi:10.1088/0264-9381/28/9/094013
[arXiv:1012.0908 [gr-qc]].
%747 citations counted in INSPIRE as of 07 Nov 2023

%57
%\cite{Reitze:2019iox}
\bibitem{Reitze:2019iox}
D.~Reitze, R.~X.~Adhikari, S.~Ballmer, B.~Barish, L.~Barsotti, G.~Billingsley, D.~A.~Brown, Y.~Chen, D.~Coyne and R.~Eisenstein, \textit{et al.}
Cosmic Explorer: The U.S. Contribution to Gravitational-Wave Astronomy beyond LIGO,
Bull. Am. Astron. Soc. \textbf{51}, 035 (2019)
[arXiv:1907.04833 [astro-ph.IM]].
%797 citations counted in INSPIRE as of 24 Nov 2023



%59
%\cite{Liu:2023qap}
\bibitem{Liu:2023qap}
C.~Liu, W.~H.~Ruan and Z.~K.~Guo, Confusion noise from Galactic binaries for Taiji,
Phys. Rev. D \textbf{107}, 064021 (2023)
%doi:10.1103/PhysRevD.107.064021
[arXiv:2301.02821 [astro-ph.IM]].
%5 citations counted in INSPIRE as of 07 Nov 2023

%60
%\cite{TianQin:2020hid}
\bibitem{TianQin:2020hid}
J.~Mei \textit{et al.} [TianQin],
The TianQin project: current progress on science and technology,
PTEP \textbf{2021}, 05A107 (2021)
%doi:10.1093/ptep/ptaa114
[arXiv:2008.10332 [gr-qc]].
%137 citations counted in INSPIRE as of 07 Nov 2023

%\cite{LISA:2017pwj}
\bibitem{LISA:2017pwj}
P.~Amaro-Seoane \textit{et al.} [LISA],
Laser Interferometer Space Antenna,
[arXiv:1702.00786 [astro-ph.IM]].
%2485 citations counted in INSPIRE as of 19 Sep 2023

%61
%\cite{Planck:2018vyg}
\bibitem{Planck:2018vyg}
N.~Aghanim \textit{et al.} [Planck],
Planck 2018 results. VI. Cosmological parameters,
Astron. Astrophys. \textbf{641}, A6 (2020)
%[erratum: Astron. Astrophys. \textbf{652}, C4 (2021)]
%doi:10.1051/0004-6361/201833910
[arXiv:1807.06209 [astro-ph.CO]].
%11918 citations counted in INSPIRE as of 30 Oct 2023

%\cite{Zhang:2019ylr}
\bibitem{Zhang:2019ylr}
X.~Zhang,
Gravitational wave standard sirens and cosmological parameter measurement,
Sci. China Phys. Mech. Astron. \textbf{62}, no.11, 110431 (2019)
%doi:10.1007/s11433-019-9445-7
[arXiv:1905.11122 [astro-ph.CO]].
%48 citations counted in INSPIRE as of 05 Feb 2024

%\cite{Poisson:1995ef}
\bibitem{Poisson:1995ef}
E.~Poisson and C.~M.~Will,
Gravitational waves from inspiraling compact binaries: Parameter estimation using second postNewtonian wave forms,
Phys. Rev. D \textbf{52}, 848-855 (1995)
%doi:10.1103/PhysRevD.52.848
[arXiv:gr-qc/9502040 [gr-qc]].
%412 citations counted in INSPIRE as of 16 Oct 2023

%63
%\cite{Finn:1992xs}
\bibitem{Finn:1992xs}
L.~S.~Finn and D.~F.~Chernoff,
Observing binary inspiral in gravitational radiation: One interferometer,
Phys. Rev. D \textbf{47}, 2198-2219 (1993)
%doi:10.1103/PhysRevD.47.2198
[arXiv:gr-qc/9301003 [gr-qc]].
%648 citations counted in INSPIRE as of 16 Oct 2023

%64
%\cite{Cutler:1994ys}
\bibitem{Cutler:1994ys}
C.~Cutler and E.~E.~Flanagan,
Gravitational waves from merging compact binaries: How accurately can one extract the binary's parameters from the inspiral wave form?,
Phys. Rev. D \textbf{49}, 2658-2697 (1994)
%doi:10.1103/PhysRevD.49.2658
[arXiv:gr-qc/9402014 [gr-qc]].
%1297 citations counted in INSPIRE as of 16 Oct 2023

%65
%\cite{Berti:2004bd}
\bibitem{Berti:2004bd}
E.~Berti, A.~Buonanno and C.~M.~Will,
Estimating spinning binary parameters and testing alternative theories of gravity with LISA,
Phys. Rev. D \textbf{71}, 084025 (2005)
%doi:10.1103/PhysRevD.71.084025
[arXiv:gr-qc/0411129 [gr-qc]].
%355 citations counted in INSPIRE as of 01 Jan 2024

%66
%\cite{Srivastava:2022slt}
\bibitem{Srivastava:2022slt}
V.~Srivastava, D.~Davis, K.~Kuns, P.~Landry, S.~Ballmer, M.~Evans, E.~D.~Hall, J.~Read and B.~S.~Sathyaprakash,
Science-driven Tunable Design of Cosmic Explorer Detectors,
Astrophys. J. \textbf{931}, 22 (2022)
%doi:10.3847/1538-4357/ac5f04
[arXiv:2201.10668 [gr-qc]].
%33 citations counted in INSPIRE as of 04 Nov 2023

%67
%\cite{Li:2023gtu}
\bibitem{Li:2023gtu}
T.~N.~Li, S.~J.~Jin, H.~L.~Li, J.~F.~Zhang and X.~Zhang,
Prospects for probing the interaction between dark energy and dark matter using gravitational-wave dark sirens with neutron star tidal deformation,
[arXiv:2310.15879 [astro-ph.CO]].
%2 citations counted in INSPIRE as of 31 Dec 2023

%\cite{Han:2023exn}
\bibitem{Han:2023exn}
T.~Han, S.~J.~Jin, J.~F.~Zhang and X.~Zhang,
A comprehensive forecast for cosmological parameter estimation using joint observations of gravitational-wave standard sirens and short $\gamma$-ray bursts,
[arXiv:2309.14965 [astro-ph.CO]].
%3 citations counted in INSPIRE as of 12 Jan 2024

%\cite{Wang:2021srv}
\bibitem{Wang:2021srv}
L.~F.~Wang, S.~J.~Jin, J.~F.~Zhang and X.~Zhang,
Forecast for cosmological parameter estimation with gravitational-wave standard sirens from the LISA-Taiji network,
Sci. China Phys. Mech. Astron. \textbf{65}, 210411 (2022)
%doi:10.1007/s11433-021-1736-6
[arXiv:2101.11882 [gr-qc]].
%36 citations counted in INSPIRE as of 21 Dec 2023

%68
%\cite{Klein:2015hvg}
\bibitem{Klein:2015hvg}
A.~Klein, E.~Barausse, A.~Sesana, A.~Petiteau, E.~Berti, S.~Babak, J.~Gair, S.~Aoudia, I.~Hinder and F.~Ohme, \textit{et al.}
Science with the space-based interferometer eLISA: Supermassive black hole binaries,
Phys. Rev. D \textbf{93}, 024003 (2016)
%doi:10.1103/PhysRevD.93.024003
[arXiv:1511.05581 [gr-qc]].
%394 citations counted in INSPIRE as of 21 Dec 2023


%end
\end{thebibliography}
\end{document}